\documentclass{aa}  
\usepackage{amsmath}
\usepackage{graphicx}
\usepackage{txfonts}
\usepackage{natbib}
\usepackage{color}

\bibpunct{(}{)}{;}{a}{}{,} 
\usepackage{hyperref}
\hypersetup{
        colorlinks = true,
        linkcolor = blue,
        anchorcolor = red,
        citecolor = blue,
        filecolor = red,
        pagecolor = red,
        urlcolor = red
} 

\def\ie{{\it i.e.,}\,}
\def\eg{{\it e.g.,}\,}

\def\la{\hbox{\raise.5ex\hbox{$<$} 
    \kern-1.1em\lower.5ex\hbox{$\sim$}}} 
\def\ga{\hbox{\raise.5ex\hbox{$>$} 
    \kern-1.1em\lower.5ex\hbox{$\sim$}}}

\newcommand{\dgr}{\mbox{$^\circ$}}           
\newcommand{\Msun}{\mbox{M$_\odot$\,}}         
\newcommand{\Lsun}{\mbox{$L_\odot$}}         
\newcommand{\cm}{\mbox{\ cm}}                
\newcommand{\s}{\mbox{\ s}}                  
\newcommand{\cms}{\mbox{\ cm s${}^{-1}$}}    
\newcommand{\cmss}{\mbox{\ cm s${}^{-2}$}}    
\newcommand{\radss}{\mbox{\ rad$^2 \s^{-2}$\,}}  

\begin{document}
\bibliographystyle{aa}

\title{Rayleigh-Taylor finger instability mixing in hydrodynamic
       shell convection models}

   \author{M. Moc\'ak \inst{1}
           \and E. M\"uller \inst{2}}

   \institute{Institut d'Astronomie et d'Astrophysique, Universit\'e Libre de
              Bruxelles, CP 226, 1050 Brussels, Belgium\\
              \email{mmocak@ulb.ac.be}         
          \and Max-Planck-Institut f\"ur Astrophysik,
              Postfach 1312, 85741 Garching, Germany\\
              \email{ewald@mpa-garching.mpg.de}}

   \date{Received  ........................... }

 
  \abstract
   {Mixing processes in stars driven by composition gradients as a
     result of the Rayleigh-Taylor instability are not anticipated.
     They are supported only by hydrodynamic studies of stellar 
     convection. We find that such mixing occurs below the bottom edge
     of convection zones in our multidimensional hydrodynamic 
     shell convection models.  It operates at interfaces created by
     off-center nuclear burning, where less dense gas with higher mean
     molecular weight is located above denser gas with a lower mean
     molecular weight.}
   {We discuss the mixing under various conditions with hydrodynamic 
     convection models based on stellar evolutionary calculations of
     the core helium flash in a 1.25 \Msun star, the core carbon flash
     in a 9.3\,\Msun star, and of oxygen burning shell in a star with
     a mass of 23\,\Msun.}
   {We simulate the hydrodynamic behavior of shell convection during
     various phases of stellar evolution with the Eulerian
     hydrodynamics code HERAKLES in two and three spatial dimensions.
     Initial models for this purpose are obtained by state-of-the-art
     stellar evolutionary codes, namely GARSTEC, STAREVOL, and TYCHO
     for the core helium flash, core carbon flash, and oxygen shell
     burning, respectively. Most of our analysis is performed for
     two-dimensional hydrodynamic models of shell convection during the
     core helium flash at its peak covering approximately 250
     convective turnover timescales or 1.4 days of stellar evolution.}
   {The mixing manifests itself in the form of overdense and cold
     fingers enriched with matter of higher mean molecular weight,
     originating from density fluctuations at the lower boundary of
     the convective shell, and ``shooting'' down into the core. The fingers
     are neither produced by overshooting from the convection zone nor
     by thermohaline mixing. Instead, they result from the
     Rayleigh-Taylor instability at the lower convection zone boundary
     due to a negative mean molecular weight gradient (molecular
     weight decreasing in the direction of gravity). They do not
     appear when the mean molecular weight gradient is positive. We
     call this process Rayleigh-Taylor finger instability mixing or
     RTFI mixing for short.}
   {}


 \titlerunning{Rayleigh-Taylor finger instability mixing}

   \maketitle
%

\section{Introduction}
\label{sect:intro}

The hydrodynamic approach to model certain phases of stellar
evolution, like \eg a nuclear core flash, by numerically solving the
Euler or Navier-Stokes equations is essentially built upon first
principles in physics, and thus (almost) parameter-free. This approach
is advantageous compared to (1D) stellar evolutionary calculations
when modelling phenomena related to flow instabilities which are
inherently multidimensional in nature. One of these instabilities is
the Rayleigh-Taylor (RT) instability \citep{Rayleigh1883, Taylor1949},
caused by the buoyancy force which acts on fluid/gas elements (blobs)
of matter that are over-dense or over-light with respect to the matter
surrounding them.

One consequence of the RT instability in stellar evolution is the
occurrence of convection \citep{KipWeigert1990, CoxGiuli2008}, whose
role for stellar evolution is even nowadays still mainly known from 1D
hydrostatic stellar evolutionary calculations. However, convection is
a genuine multidimensional phenomenon, \ie its properties can be
reliably inferred only on the basis of multidimensional hydrodynamic
simulations. To this end several hydrodynamic studies of shell
convection in the cores of stars (hence sometimes called core
convection) have been performed in recent years studies, with
interesting conclusions \citep{Deupree1996, Dearborn2006, Herwig2006,
  MeakinArnett2007, Mocak2009}.  These simulations also considered the
effects of composition gradients, but without investigating their
influence in much detail, except for the work of
\citet{MeakinArnett2007} highlighting the importance of turbulent
entrainment \citep{Fernando1991} at the boundaries of convection
zones.

Composition gradients play an important role during the evolution of
stars of all masses. They can instigate double-diffusive instabilities
\citep{Kippenhahn1980, Turner1985} \ie ``salt-fingering'' or the
occurrence of vibrational overstability known as semi-convection
\citep[;and references therein]{Grossman1996} and can trigger
gravitational settling \citep{Stancliffe2008} thereby influence the
evolution of a star. Chemical gradients are also likely responsible
that a star becomes a red giant \citep{Stancliffe2009}, and they lead
to deep mixing in envelopes of low-mass red giants
\citep{Eggleton2006}.

Stellar layers with a negative mean molecular weight gradient (\ie the
molecular weight $\mu$ decreases in the direction of gravity) can
develop a shell where thermohaline mixing occurs
\citep{Kippenhahn1980}, which is driven by a double-diffusive
instability.  Thermohaline mixing can be regarded as a diffusion
process where blobs of gas with higher molecular weight sink down,
while other material rises and mixes with neighboring gas, eventually
providing further rising and sinking elements.  Such a situation is
encountered, \eg during the off-center ignition of helium in the helium
core during the core helium flash, where a convective shell forms
above the ignition point which is continuously enriched by nuclear
ashes with a higher molecular weight (mainly $^{12}$C). This creates a
negative mean molecular weight gradient at the base of the convection
zone, which should cause thermohaline mixing within a thermal
timescale \citep{Kippenhahn1980}. Thermohaline mixing is supposed to
manifest itself in the form of finger-like structures
\citep{Turner1985}, which are indeed observed, \eg in oceanography
\citet{Williams1974, Gargett1982}. On the other hand, results from
experimental chemo-hydrodynamics often reveal instabilities that have
a finger-like structure as well, but which are not regarded as
thermohaline mixing. Well known in this context \citep{AnneDeWit2004,
AnneDeWit2008} is viscous fingering  where a less viscous fluid is 
displacing a more viscous one in porous media (Saffamn-Taylor instability), 
and density fingering driven by a Rayleigh-Taylor instability in systems, 
where a denser solution resides above a lighter one.

In the following we discuss a mixing process characterized by sinking
dense blobs of gas that carry matter of higher mean molecular weight
$\mu$ rapidly and presumably adiabatically downwards (in the direction
of gravity) into matter of lower $\mu$, leaving behind trails which
resemble finger-like structures.  This process that has not been
anticipated by stellar evolutionary calculations is neither the result
of thermohaline mixing, nor viscous fingering, nor density
fingering. The mixing process operates below the base of convection
zones where the mean molecular weight gradient is negative, \ie where
``lighter'' matter resides above ``denser'' matter. Hence, it does not
occur below convection zones where the mean molecular weight gradient
is positive.  As it is caused by a Rayleigh-Taylor instability, we
call it Rayleigh-Taylor Finger Instability mixing, or RTFI mixing for
short.  Intrinsic properties of the finger-like structures inferred
mainly from our hydrodynamic simulations of shell convection during
the core helium flash rule out an explanation within the framework of
thermohaline mixing as the finger-like structures are not hotter, but
colder than the ambient medium. This fact also supports findings of
\citet{Kippenhahn1980}, who estimated based on calculations of
\citet{Thomas1967} for the non-central core helium flash of a 1.3
\Msun star that the timescale for thermohaline mixing (the time it
takes diffusion to double the geometrical thickness of the partially
mixed region below the convection zone) is of the order 10$^5\,$yrs,
which is much too slow to be seen in our hydrodynamic simulations. It
is neither expected that the timescale will drastically change
(decrease) with improved stellar models nor during the flash as the
conditions are in general very similar \citep{SweigertGross1978}.

\begin{figure} 
\includegraphics[width=0.95\hsize]{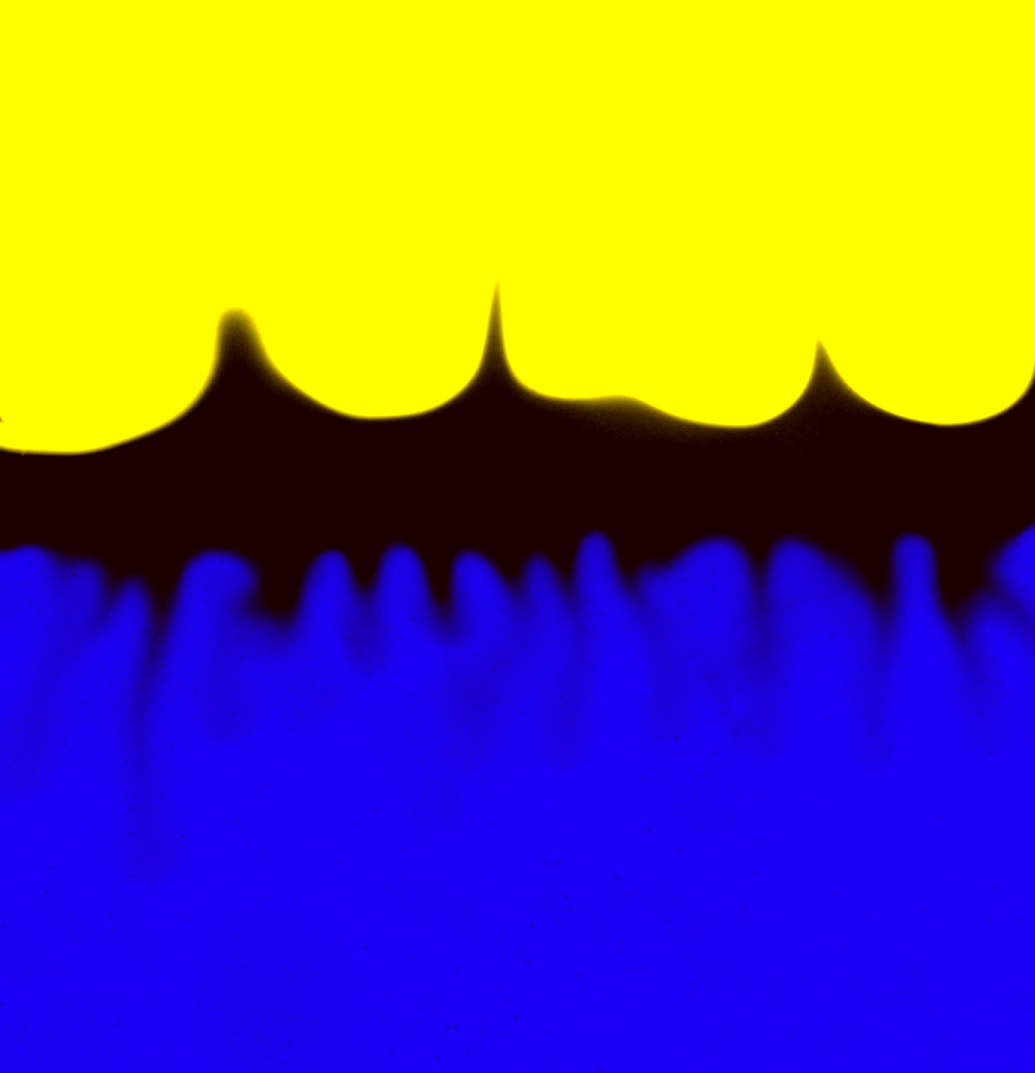}
\vspace{1.2cm}
\caption{Buoyancy-driven deformation of an interface between a less
  dense solution of HCl (yellow) residing above a denser solution of
  NaOH (blue) in an external gravitational field (pointing downwards in
  the figure). The field of view is 3 by 3\,cm, and the snapshot is
  taken few minutes after contact \citep{Almarcha2010}.}
\label{fig.expfing}
\end{figure}

Contrary to the thermohaline instability, a dynamic mixing process
like RTFI mixing is unknown in experimental chemo-hydrodynamics.  It
is unexpected as matter is mixed from a less dense layer into a denser
one.  However, similarly looking buoyancy-driven instabilities were
indeed found in the laboratory at the interface between a solution of
lighter hydrochloric acid (HCl) residing above a heavier solution of
sodium hydroxide (NaOH) oriented vertically in the gravitional field
by \citet{Zalts2008}. The reaction of these two solutions at their
common interface leads to the production of salt (NaCl), and to
over-dense (with respect to the horizontal average) fingers of NaCl
extending into the NaOH solution as shown in a snapshot from the
experiment of \citet{AnneDeWit2008} or \citet{Almarcha2010} in
Fig.\,\ref{fig.expfing}.  Such systems resemble principle similarities
with the stellar shell convection models where lighter gas of higher
mean molecular weight is residing above denser gas of lower mean
molecular weight. Proposed explanations for this kind of laboratory
mixing range from amplified local density fluctuations due to salt
production, a local density decrease due to a temperature rise up to
differences in the mass diffusion coefficients of the reacting species
\citep{Zalts2008}.  Recent work seems to favor the latter one
\citep{Almarcha2010}.  However, mass diffusion in stars is too slow to
be responsible for the mixing observed in our simulations (see \eg
\citet{Michaud2008}).

\begin{table} 
\caption[]{Some characteristic properties of our initial models: total
  mass $M$, stellar population, metal content $Z$, mass $M_{m}$, outer
  (inner) radius $R_{m}$, and nuclear energy production rate $L_{m}$,
  respectively.}
\vspace{0.5cm}
\begin{tabular}{l|lcllll} 
Model & $M$  & Pop. & $Z$ & $M_{m}$  & $R_{m}$    & 
$L_{m}$    \\ 
      & $[\Msun]$ &      &     & $[\Msun]$ & $[10^9\cm]$ &
$[10^9\Lsun]$ \\
\hline 
M  & $1.25$ & I & $~~~0.02$ & $0.45$ & $1.2(0.2)$ & $\sim 1$  \\
L  & $9.3 $ & I & $~~~0.02$ & $0.94$ & $1.0(0.3)$  & $\sim 0.01$ \\
O  & $23. $ & I & $\sim 0.02$ & $2.67$ & $1.0(0.3)$  & $\sim 1000$  \\
\end{tabular} 
\label{imodtab} 
\end{table} 

\begin{figure} 
\includegraphics[width=0.99\hsize]{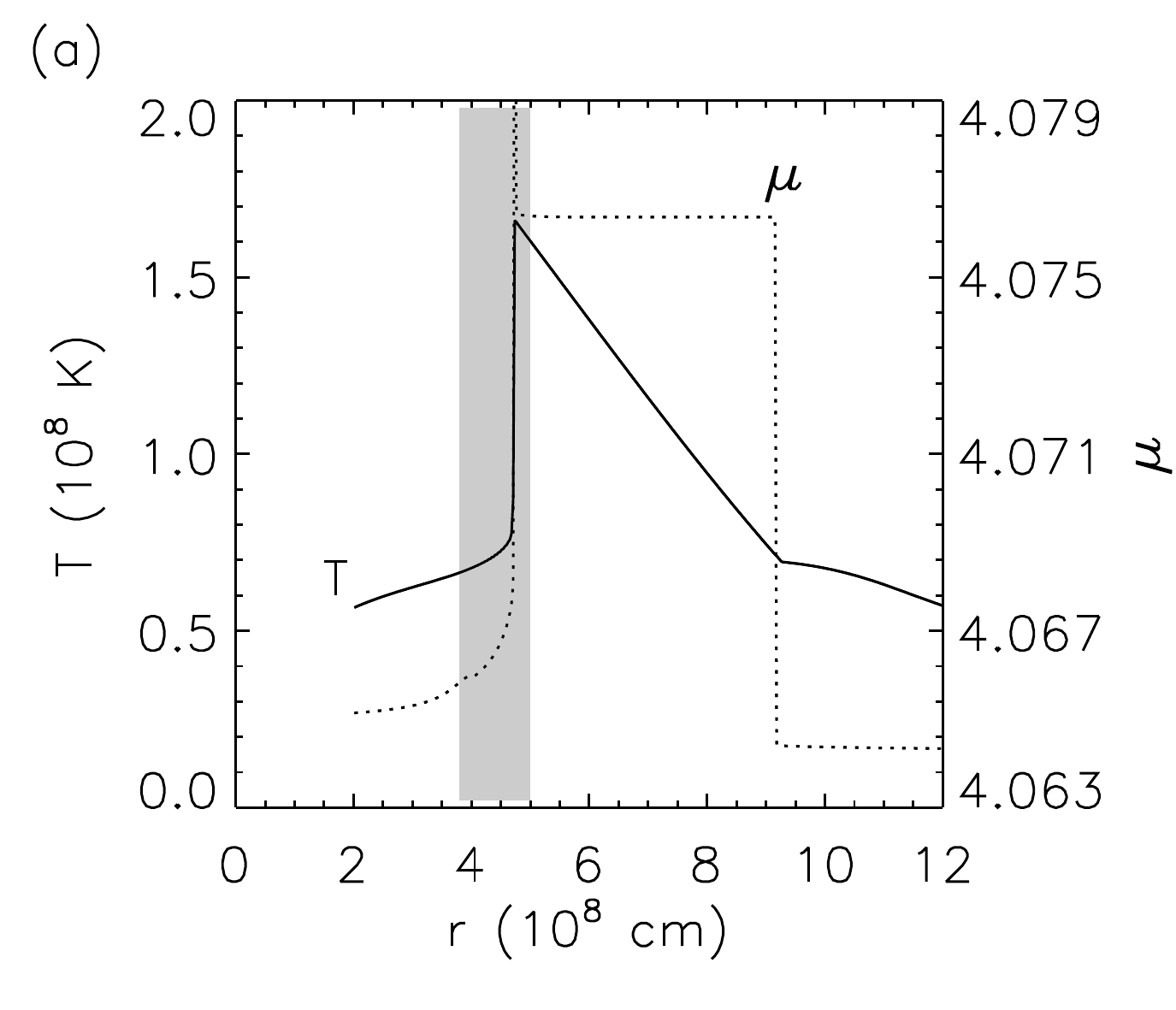}
\vspace{-0.5cm}
\includegraphics[width=0.99\hsize]{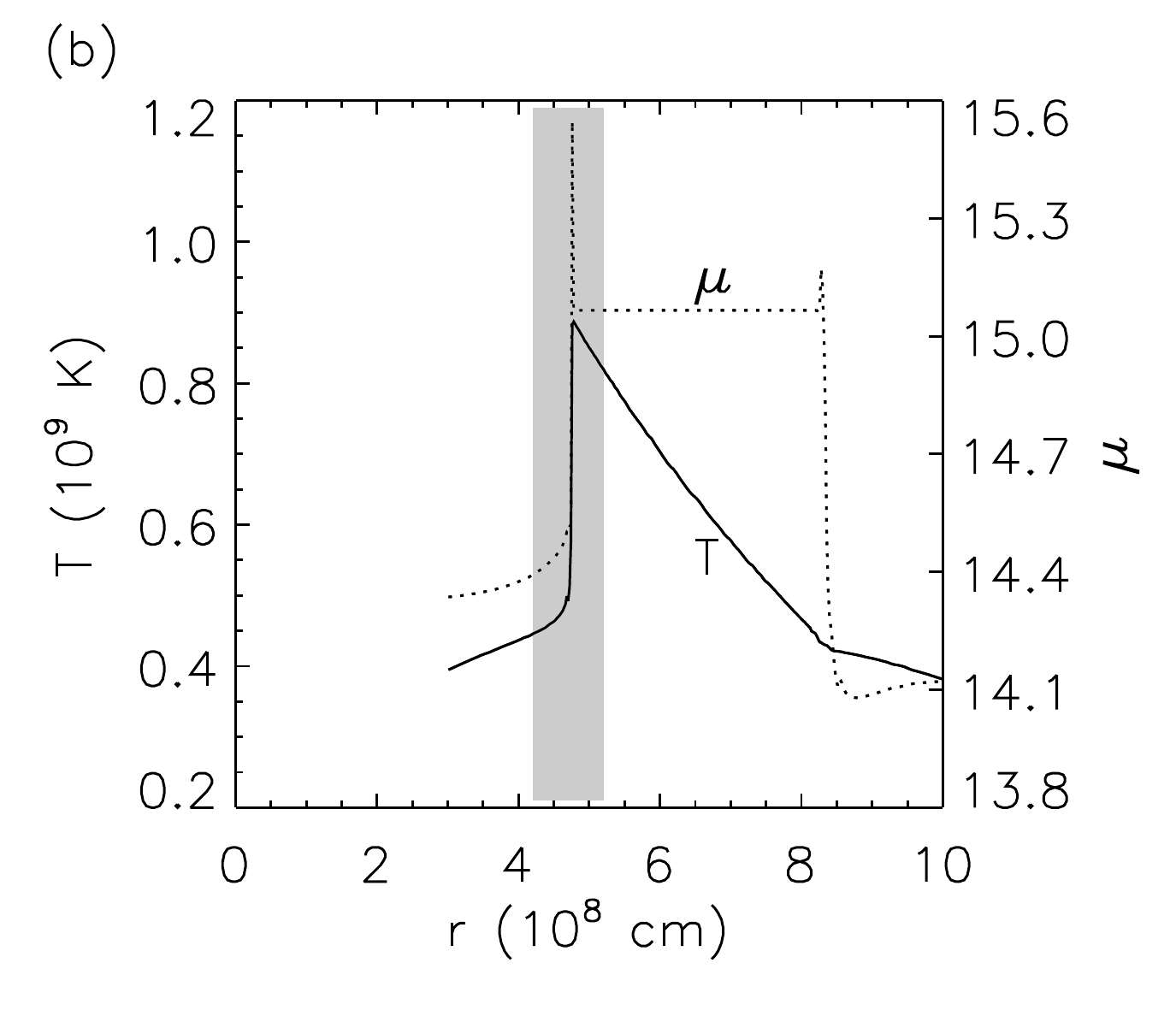}
\vspace{-0.5cm}
\includegraphics[width=0.99\hsize]{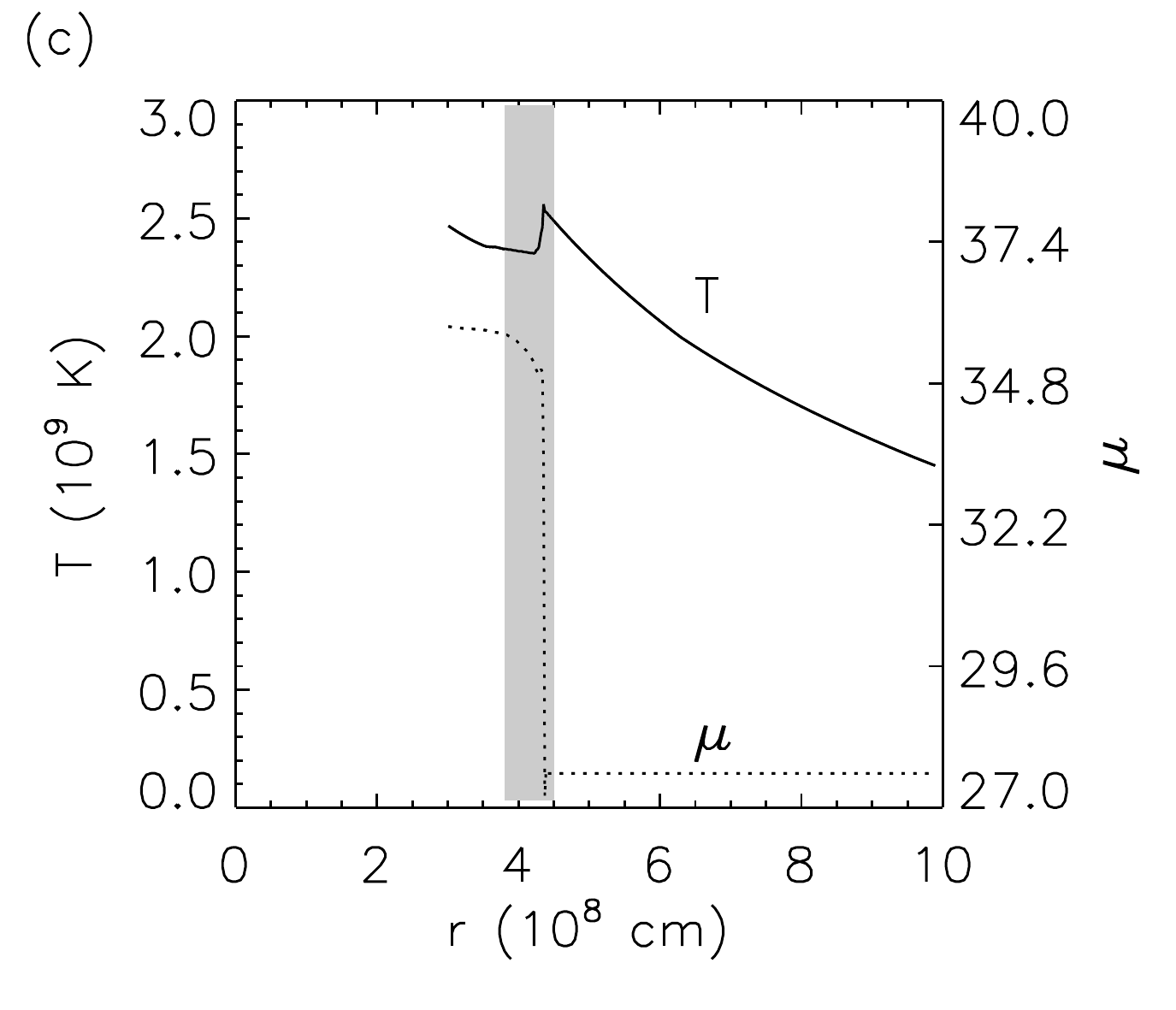}
%
\caption{Temperature $T$ (solid) and mean molecular weight $\mu$
  (dotted) as a function of radius for the initial core helium flash
  model (a), the core carbon flash model (b), and the oxygen shell
  burning model (c), respectively . In each of these panels the region
  of interest below the base of the convection zone is highlighted by
  the shaded vertical strip. 
 }
\label{fig.inimod}
\end{figure} 

The paper is organized as follows. In Sect.\,\ref{sect:inidat} we
discuss the input models for our simulations. Our hydrodynamic code
and the hydrodynamic shell convection models together with the local
and temporal properties of the RTFI mixing are described in
Sect.\,\ref{sect:hydrosim}. A possible theoretical explanation of RTFI
mixing is discussed in Sect.\,\ref{sect:locstab}, while the breaking
of dynamic stability in stars is analyzed in
Sect,\,\ref{sect:disc}. Finally, a summary of our findings is given in
Sect.\,\ref{sect:sum}.

\section{Initial Data}
\label{sect:inidat}

We used three initial models for our simulations (Tab.\,\ref{imodtab},
Fig.\,\ref{fig.inimod}), which correspond to the helium core of a 1.25
\Msun star during the peak (maximum core luminosity) of the core
helium flash computed with the GARSTEC code \citep{WeissSchlattl2000,
  WeissSchlattl2007}, the carbon-oxygen (C-O) core of a 9.3 \Msun star
at peak of the core carbon flash computed with the STAREVOL code
\citep{Siess2006}, and the core of a 23 \Msun star during oxygen shell
burning \citep{MeakinArnett2007} computed with the TYCHO code
\citep{Young2005}.  The thermodynamic structure of all three initial
models is qualitatively similar exhibiting of an off-center
temperature maximum due to nuclear burning in a partially electron
degenerate stellar core. Directly outside this temperature maximum the
inner boundary of a convective shell is located which extends towards
larger radii.

The core helium flash occurs in low-mass stars ($0.7\,\Msun \le M
\lesssim 2.2\,\Msun$) due to an off-center ignition of helium by the
triple-$\alpha$ reaction after central hydrogen exhaustion in the
semi-degenerate helium core. The core carbon flash ensues after
central helium exhaustion and is characterized by off-center carbon
ignition in a semi-degenerate carbon core of a rather massive star
($9\,\Msun \lesssim M \le 12\,\Msun$), the dominant nuclear reactions
being $^{12}$C\,($^{12}$C, $\alpha$)$\,^{20}$Ne and
$^{12}$\,C($^{12}$C, p)$\,^{23}$Na followed by $^{16}$O\,($\alpha$,
$\gamma$)$\,^{20}$Ne.  Oxygen shell burning is typical for massive
stars ($M \gtrsim 9\,\Msun$) close to core collapse, when oxygen and
carbon burn in a convective shell above a silicon-sulfur rich
semi-degenerate core possessing an off-center temperature maximum
\citep{Arnett1994, Bazan1998, Asida2000, Meakin2006,MeakinArnett2007}.

\begin{figure*} 
\includegraphics[width=0.33\hsize]{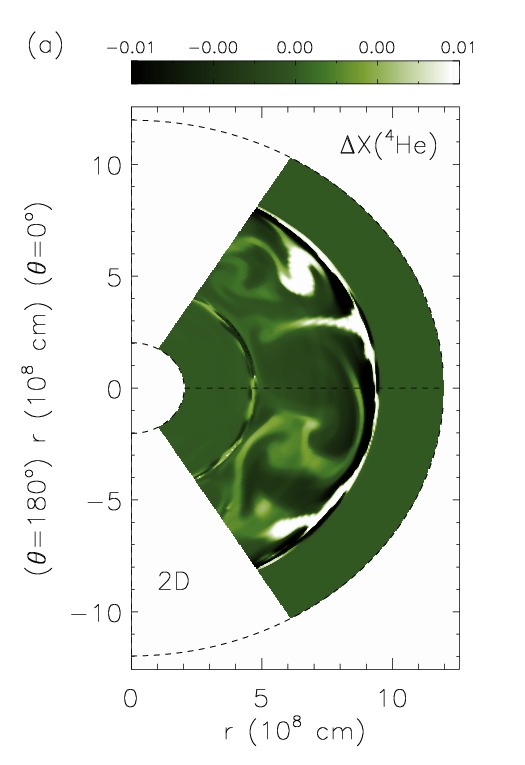}
\includegraphics[width=0.33\hsize]{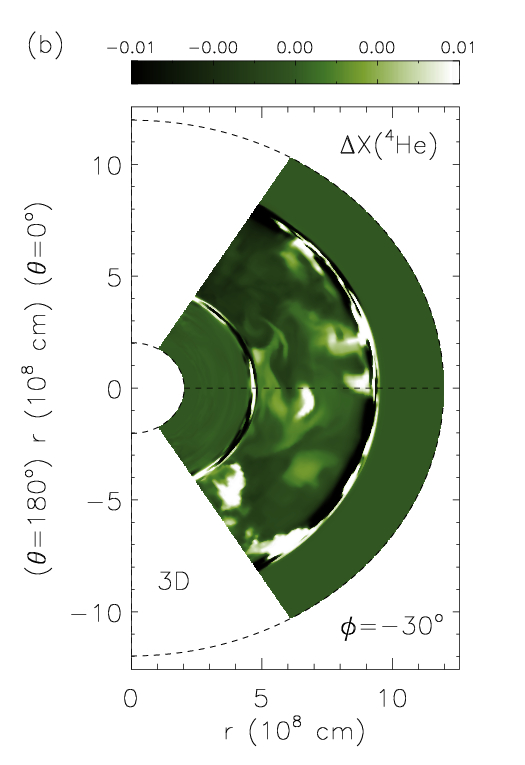}
\includegraphics[width=0.33\hsize]{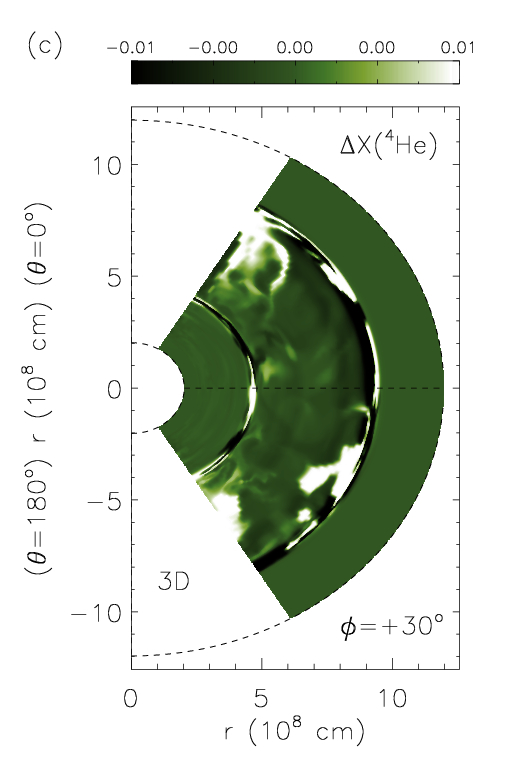}
\caption{Snapshots of the relative angular fluctuations of the helium
  mass fraction, $\Delta X(^{4}\mbox{He}) \equiv 100\times [
    X(^{4}\mbox{He}) - \langle X(^{4}\mbox{He}) \rangle_{\theta} ]
  \,/\,\langle X(^{4}\mbox{He}) \rangle_{\theta}$ for the 2D model
  hefl.2d.1 taken at $t = 4638\,$s (a), and the 3D model hefl.3d in a
  meridional plane with azimuthal angle $\phi = -30 \dgr$ (b) and
  $\phi = +30\dgr$ (c) at $t = 6000\,$s, respectively.  $\langle
  \rangle_{\theta}$ denotes the angular average at a given radius.}
\label{fig.hydhefl}
\end{figure*} 

During the core helium flash the convective shell is enriched mainly
by $^{12}$C which results in a negative mean molecular weight gradient
below its base (Fig.\,\ref{fig.inimod}\,a)
\footnote{The gradient grows for roughly 10\,000 years since the onset
  of the core helium flash.}.
The situation is similar during the core carbon flash, where the
nuclear ash from carbon burning increases the mean molecular weight
inside its convective shell relative to that of the unburned inner
carbon core (Fig.\,\ref{fig.inimod}\,b). Oxygen shell burning, which
follows oxygen core burning, causes a positive composition gradient,
\ie the mean molecular weight increases in the direction of gravity at
the bottom of the convective shell (Fig.\,\ref{fig.inimod}\,c).

\section{Hydrodynamic Simulations}
\label{sect:hydrosim}

\begin{table} 
\caption{Some properties of the 2D and 3D simulations: number of grid
  points in $r$ ($N_{r}$), $\theta$ ($N_{\theta}$), and $\phi$
  ($N_{\phi}$) direction, radial grid resolution $\Delta r$, angular
  grid resolution $\Delta \theta$ and $\Delta \phi$ in $theta$ and
  $phi$-direction, and maximum evolutionary time $t_{max}$ of the
  simulation, respectively. }
\begin{center}
\begin{tabular}{p{1.1cm}|p{0.4cm}p{0.4cm}p{0.4cm}p{0.8cm}p{0.4cm}p{0.4cm}p{0.8cm}} 
\hline
\hline
model & N$_r$ & N$_\theta$ & N$_\phi$ & $\Delta$r & $\Delta\theta$ & $\Delta
\phi$ & t \\
 & $\#$ & $\#$ & $\#$ & [$10^6$cm] & [$\dgr$] & [$\dgr$] & [10$^3$s]
\\
\hline 
hefl.2d.1 &  180 &   90 &  - & 5.55 & 1.33 &    - &  30 \\
hefl.2d.2 &  270 &  180 &  - & 3.70 & 1    &    - &  30 \\
hefl.2d.3 &  360 &  240 &  - & 2.77 & 0.75 &    - & 120 \\
hefl.3d   &  180 &   90 & 90 & 5.55 & 1.33 & 1.33 &   3 \\
cafl.2d   &  360 &  180 &  - & 1.95 & 0.5  &    - &  10 \\
oxfl.2d   &  400 &  320 &  - & 1.75 & 0.28 &    - &   1\\ 
\hline
\end{tabular} 
\end{center}
\label{tab:hydromod}
\end{table}  

The numerical simulations were performed with a modified version of
the hydrodynamic code Herakles \citep{Mocak2008}. The code employs the
PPM reconstruction scheme \citep{ColellaWoodward1984}, a Riemann
solver for real gases according to \citet{ColellaGlaz1984}, and the
consistent multi-fluid advection scheme of \citet{PlewaMueller1999}.
Self-gravity, thermal transport and nuclear burning are included in
the code.  Nuclear reaction networks are generated using a tool of
Thielemann (private communication). In Table\,\ref{tab:hydromod} we
summarize some characteristic parameters of our 2D and 3D simulations
based on the initial models given in Tab.\,\ref{imodtab}.

The core helium flash simulations were performed on an equidistant
spherical polar grid in 2D ($r, \theta$) and 3D ($r, \theta,
\phi$). While the angular grid of the axisymmetric 2D simulations
hefl.2d.2 and hefl.2d.3 covered the full angular range, \ie $0\dgr \le
\theta \le 180\dgr$, the 2D simulation hefl.2d.1 was restricted to the
angular region $30\dgr \le \theta \le 150\dgr$.  The 3D simulation
hefl.3d was performed within an angular wedge given by $30\dgr \le
\theta \le 150\dgr$ and $-60\dgr \le \phi \le +60\dgr$,
respectively. This allowed us to simulate the 3D model with a
reasonable amount of computational time using a radial and angular
resolution comparable to that of the 2D model hefl.2d.1 for several
convective turnover timescales (Fig.\,\ref{fig.hydhefl}). In all core
helium flash simulations the radial grid ranged from $r = 2\times
10^8\,$cm to $r = 1.2\times 10^9\,$cm. Boundary conditions were
reflective in all directions except for simulations hefl.2d.1 and
hefl.3d, where we utilized periodic boundary conditions in angular
direction(s) to avoid a numerical bias in case of wide angular
convective structures. Abundance changes due to nuclear burning were
described by a reaction network consisting of the four $\alpha$-nuclei
$^4$He, $^{12}$C, $^{16}$O, and $^{20}$Ne coupled by seven reactions.

\begin{figure*} 
\includegraphics[width=0.33\hsize]{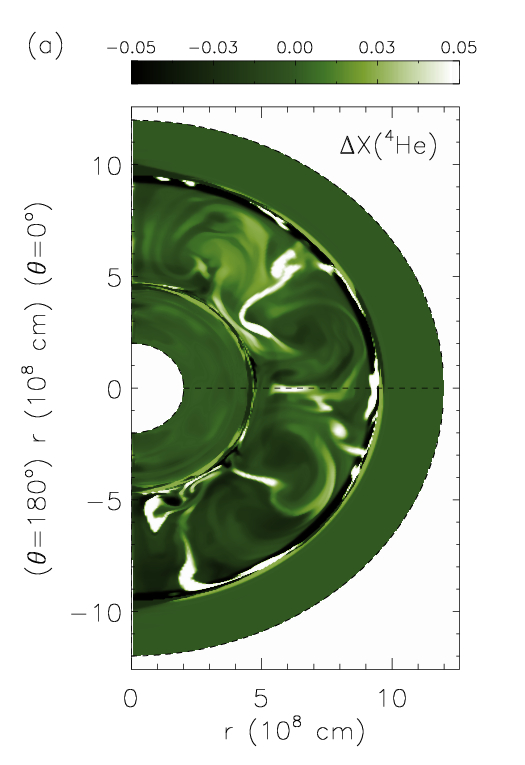}
\includegraphics[width=0.33\hsize]{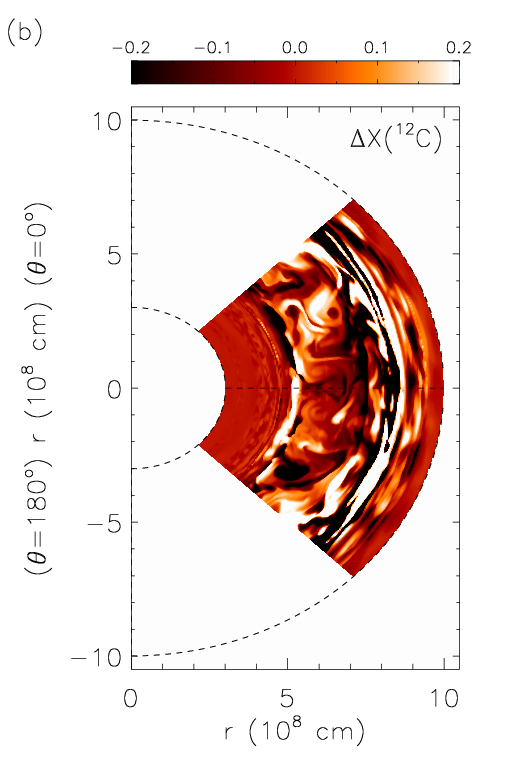}
\includegraphics[width=0.33\hsize]{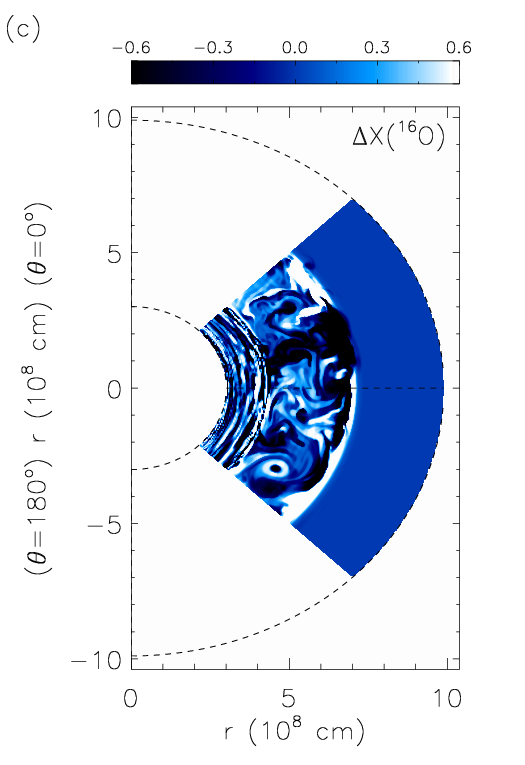}
\caption{Snapshots of the relative angular fluctuations of the mass
  fraction of $^{4}$He during the core helium flash for model
  hefl.2d.3 at $t \sim 12000\,$s (a), of $^{12}$C during the core
  carbon flash for model cafl.2d at $t \sim 682\,$s (b), and of
  $^{16}$O during oxygen shell burning for model oxfl.2d at $t \sim
  940\,$s (c), respectively. The fluctuations are defined by $\Delta
  X(^{A}{\cal N}) = \mbox{100}\times [X(^{A}{\cal N}) - \langle
    X(^{A}{\cal N}) \rangle_{\theta}] \,/\, \langle X(^{A}{\cal N})
  \rangle_{\theta}$, where $^{A}{\cal N} \in \{ ^4\mbox{He},\,
  ^{12}\mbox{C},\, ^{16}\mbox{O} \}$, and $\langle \rangle_{\theta}$
  denotes the angular average at a given radius.}
\label{fig.heflcfloxfl}
\end{figure*}

The core carbon flash simulation cafl.2d was performed on a 2D
equidistant spherical polar grid ($r, \theta$) covering a 90$\dgr$
angular wedge centered at the equator ($\theta = 90\dgr$), and a
radial grid ranging from $r = 3\times 10^8\,$cm to $r = 1\times
10^9\,$cm (Fig.\,\ref{fig.heflcfloxfl}).  Boundary conditions were
reflective in radial direction and periodic in angular direction.
Abundance changes due to nuclear burning were described by a reaction
network consisting of $^1$H, $^4$He, $^{12}$C, $^{14}$N, $^{16}$O,
$^{20}$Ne, $^{22}$Ne, $^{23}$Na, and $^{24}$Mg coupled by 17
reactions.

Oxygen shell burning simulation oxfl.2d was performed on a 2D
equidistant spherical polar grid covering a 90$\dgr$ wedge centered at
the equator, and a radial grid ranging from $r = 2\times 10^8\,$cm to
$r = 1\times 10^9\,$cm (Fig.\,\ref{fig.heflcfloxfl}). Boundary
conditions were reflective in radial direction and periodic in angular
direction. Abundance changes due to nuclear burning were described by
a reaction network consisting of neutrons, $^1$H, $^4$He, $^{12}$C,
$^{16}$O, $^{20}$Ne, $^{23}$Na, $^{24}$Mg, $^{28}$Si, $^{31}$P,
$^{32}$S, $^{34}$S, $^{35}$Cl, $^{36}$Ar, $^{38}$Ar, $^{39}$K,
$^{40}$Ca, $^{42}$Ca, $^{44}$Ti, $^{46}$Ti, $^{48}$Cr, $^{50}$Cr,
$^{52}$Fe, $^{54}$Fe, and $^{56}$Ni coupled by 76 reactions.

In the following subsections, we will briefly discuss one by one the
main properties of our hydrodynamic shell convection models of the
core helium flash, the core carbon flash, and the oxygen shell
burning.  This provides the basis for an assessment of RTFI mixing and
its importance in a wider context.

\subsection{Shell convection}
\label{subsec:coreconv}

\subsubsection{Core helium flash}
The hydrodynamic properties of shell convection during the core helium
flash are described in detail in \citet{Mocak2008, Mocak2009}.
Convection starts early ($t < 1000\,$s) and quickly extends over the
whole convectively unstable region as predicted by mixing length
theory (MLT). In axisymmetry, the convection is characterized by fast
and large circular vortices, while in 3D the convective flow is less
ordered showing slower and smaller turbulent features. The relative
fluctuations of density $\rho$, temperature $T$, and carbon $^{12}$C
with helium $^4$He in the convection zone are of the order of
10$^{-4}$, 10$^{-4}$, 10$^{-2}$ and 10$^{-5}$, respectively
(Fig.\,\ref{fig.hydhefl}). The convective velocities in our 3D model
match those predicted by MLT quite well ($|v| < 1 \times 10^{6} \cms
$), whereas the velocities in our 2D models exceed those by up to a
factor of four. Turbulent entrainment \citep{MeakinArnett2007} leads
to a growth of the width of the convection zone on a dynamic timescale
in both the 2D and 3D models due to an exchange between the potential
energy contained in the stratified layers at the boundaries of the
convection zone and the kinetic energy of the turbulent flow inside
the convection zone.

\begin{figure*} 
\includegraphics[width=0.47\hsize]{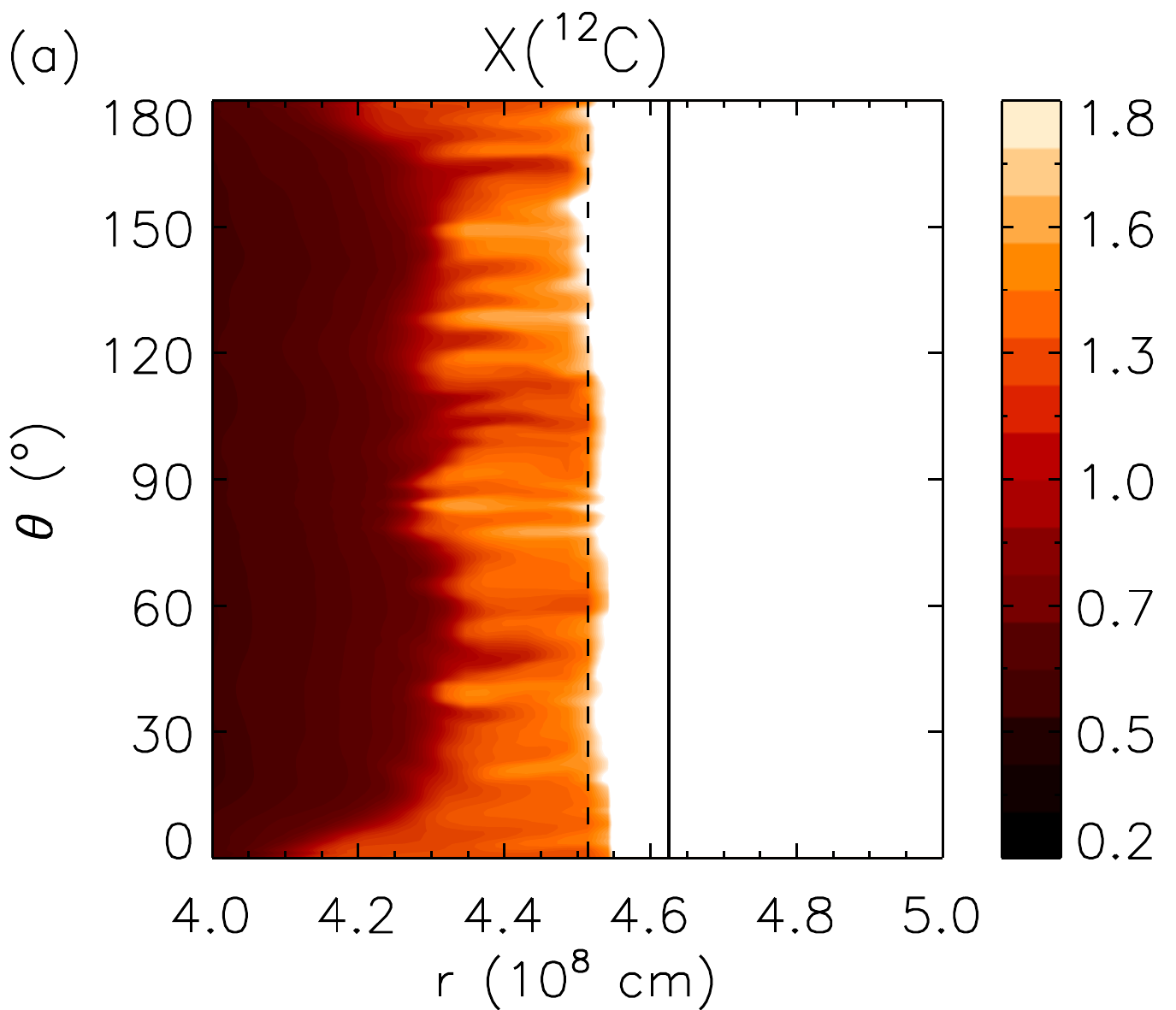}
\hfill
\includegraphics[width=0.47\hsize]{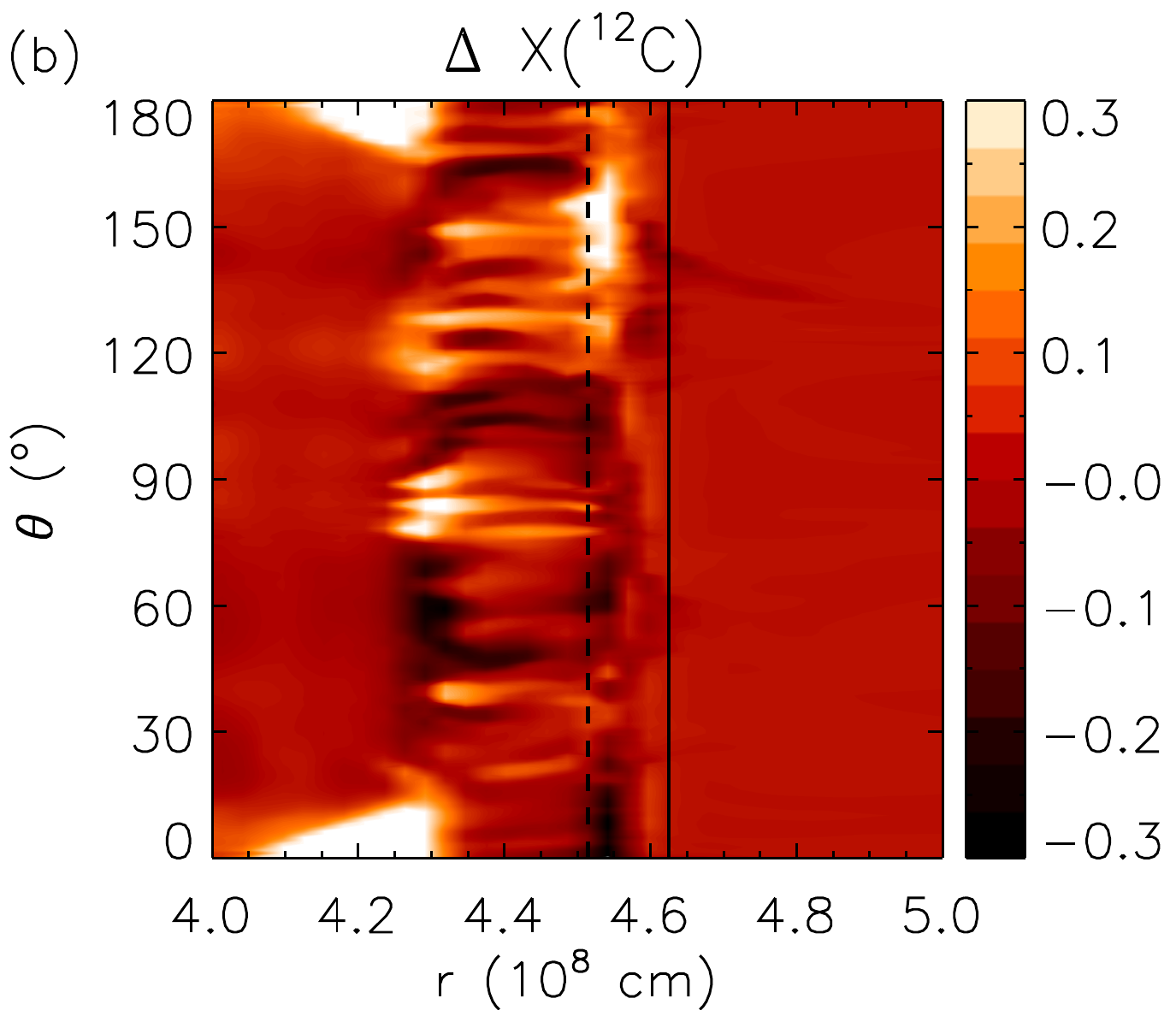}
\\
\includegraphics[width=0.47\hsize]{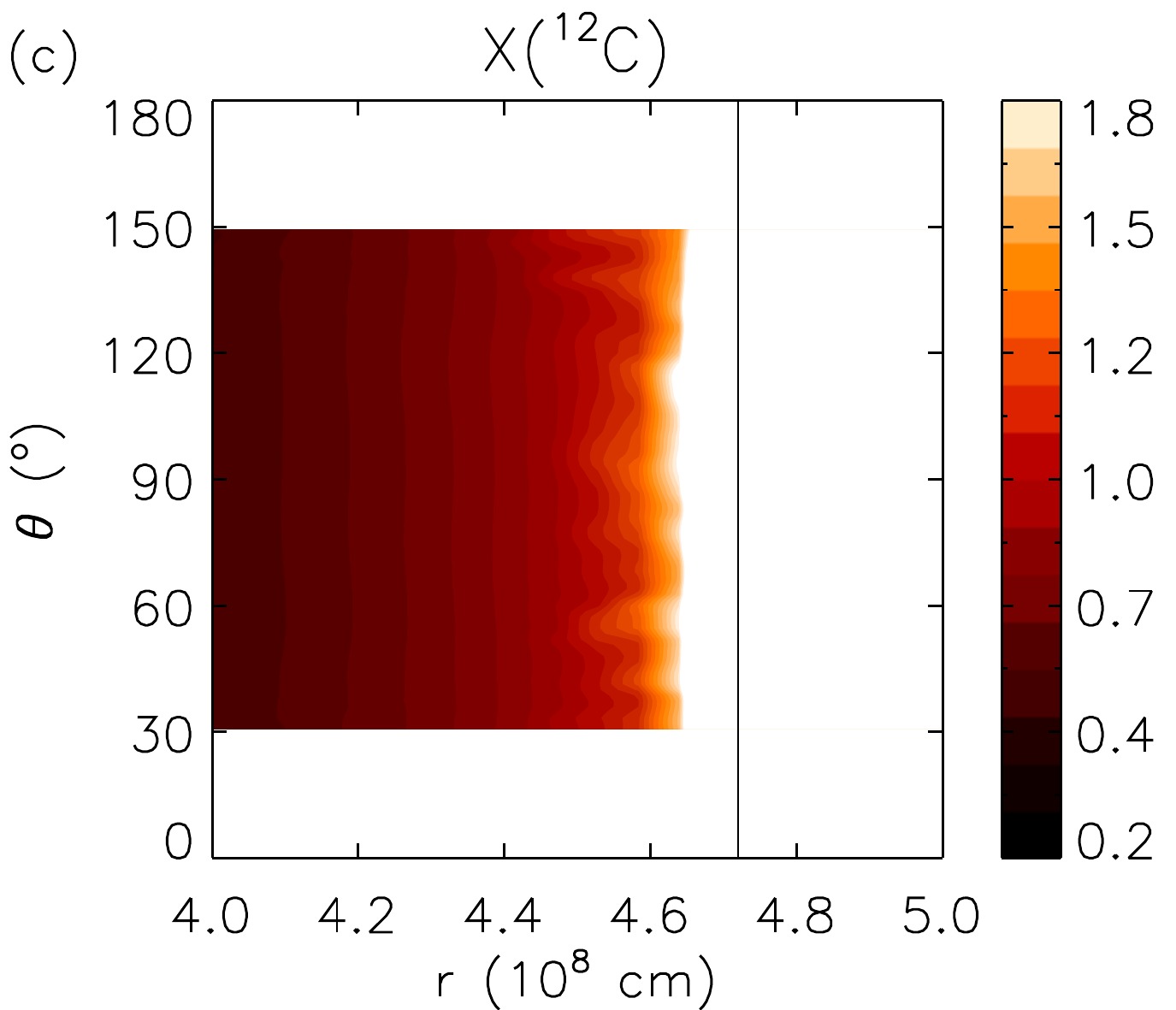}
\hfill
\includegraphics[width=0.47\hsize]{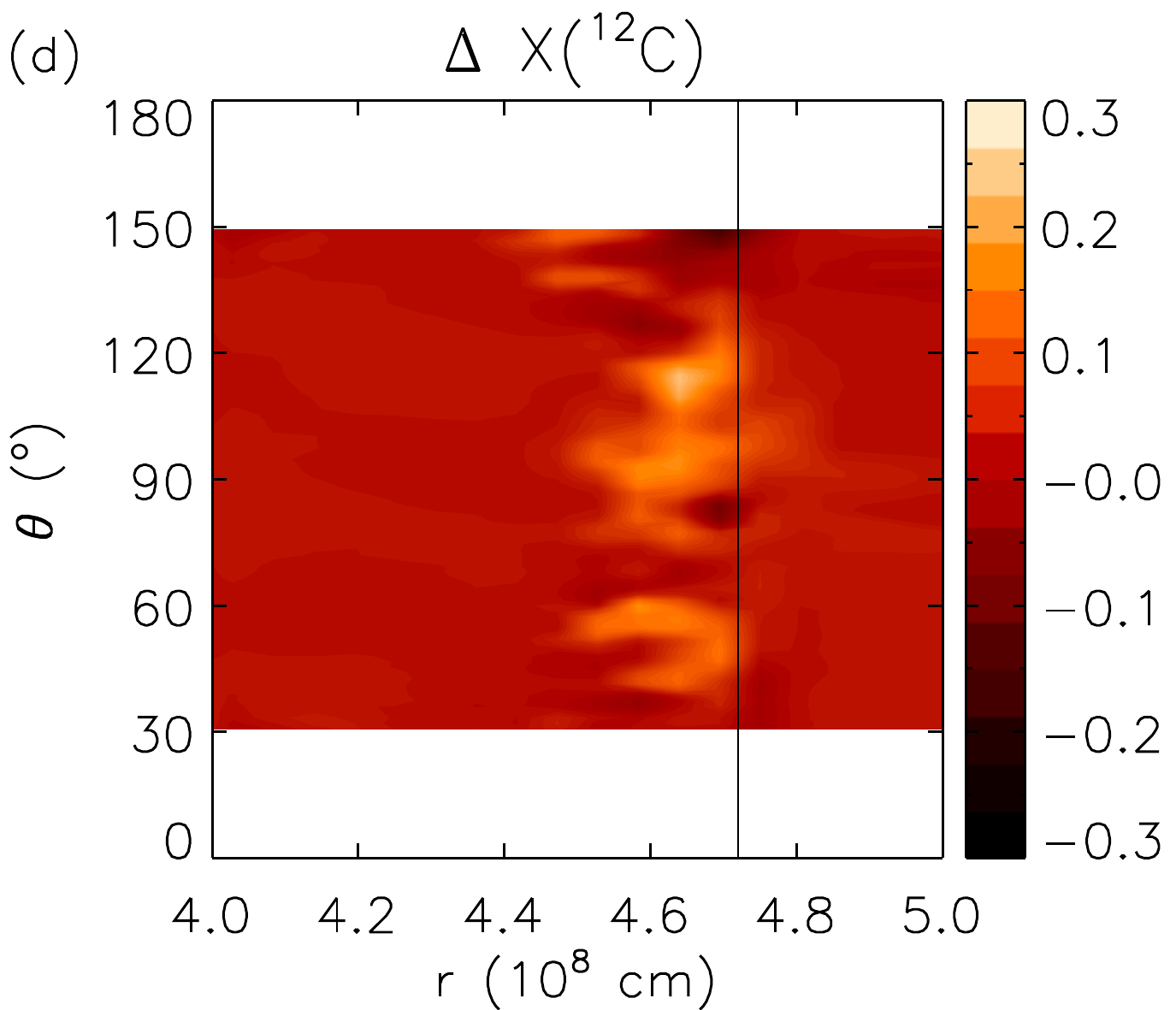}
\vspace{1.cm}
\caption{Maps of the carbon mass fraction $X(^{12}\mbox{C})$ (in units
  of $10^{-3}$) and its relative angular fluctuation $\Delta
  X(^{12}\mbox{C}) \equiv 100\times (X(^{12}\mbox{C}) - \langle
  X(^{12}\mbox{C}) \rangle_{\theta}) \,/\, \langle X(^{12}\mbox{C})
  \rangle_{\theta}$ at the bottom of the convection zone in the 2D
  model hefl.2d.3 at $t = 63430\,$s (two upper panels), and in a
  meridional plane of the 3D model hefl.3d at $t = 6000\,$s (two lower
  panels), respectively. The vertical solid line marks the bottom
  boundary of the convection zone which is equal to the position of
  $T_{max}$, and the dashed line gives the location from where RTFI
  mixing is launched. $\langle \rangle_{\theta}$ denotes the
  horizontal average at a given radius.}
\label{fig.fingcarb}  
\end{figure*} 

\begin{figure*} 
\includegraphics[width=0.47\hsize]{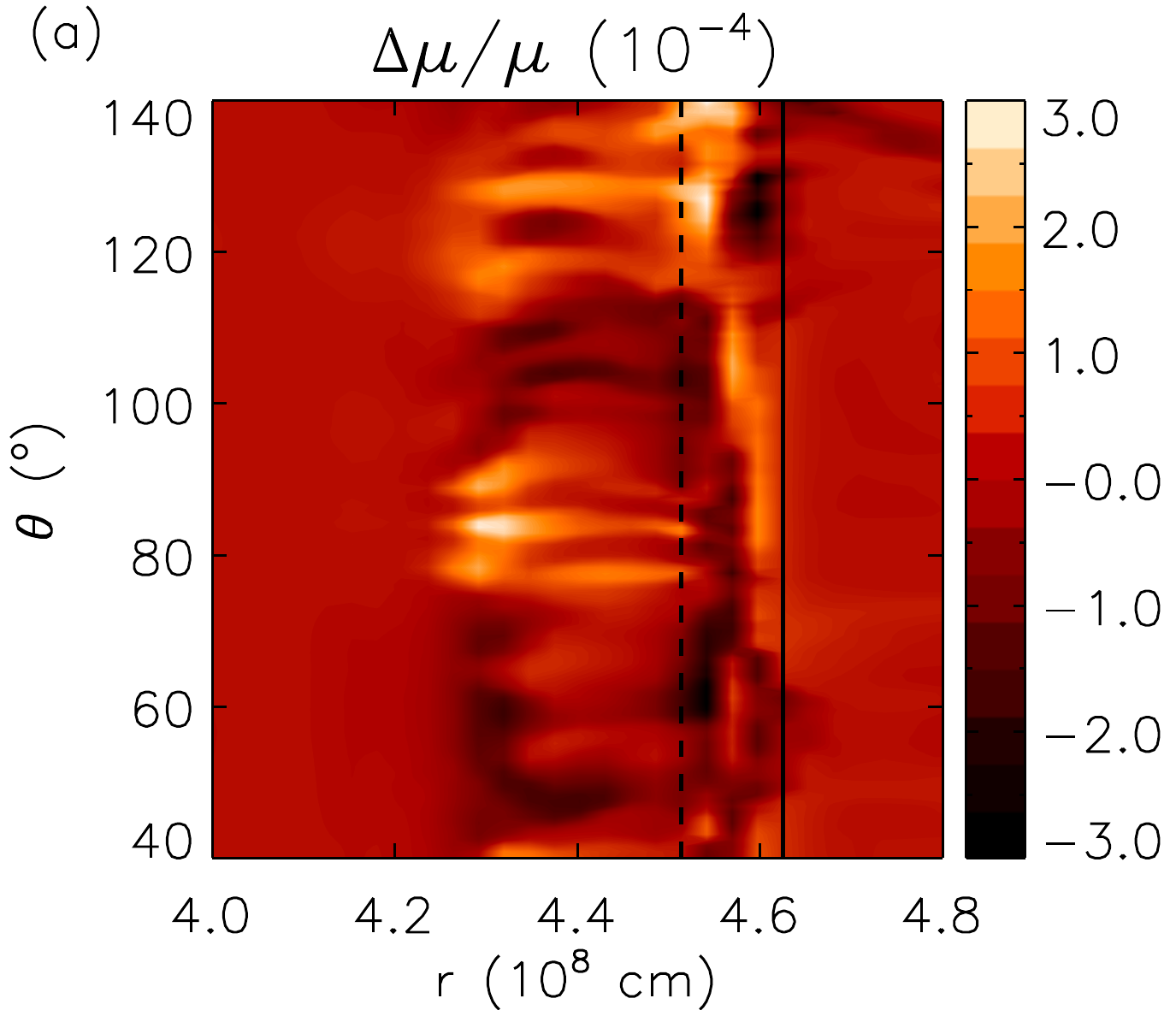}
\hfill
\includegraphics[width=0.47\hsize]{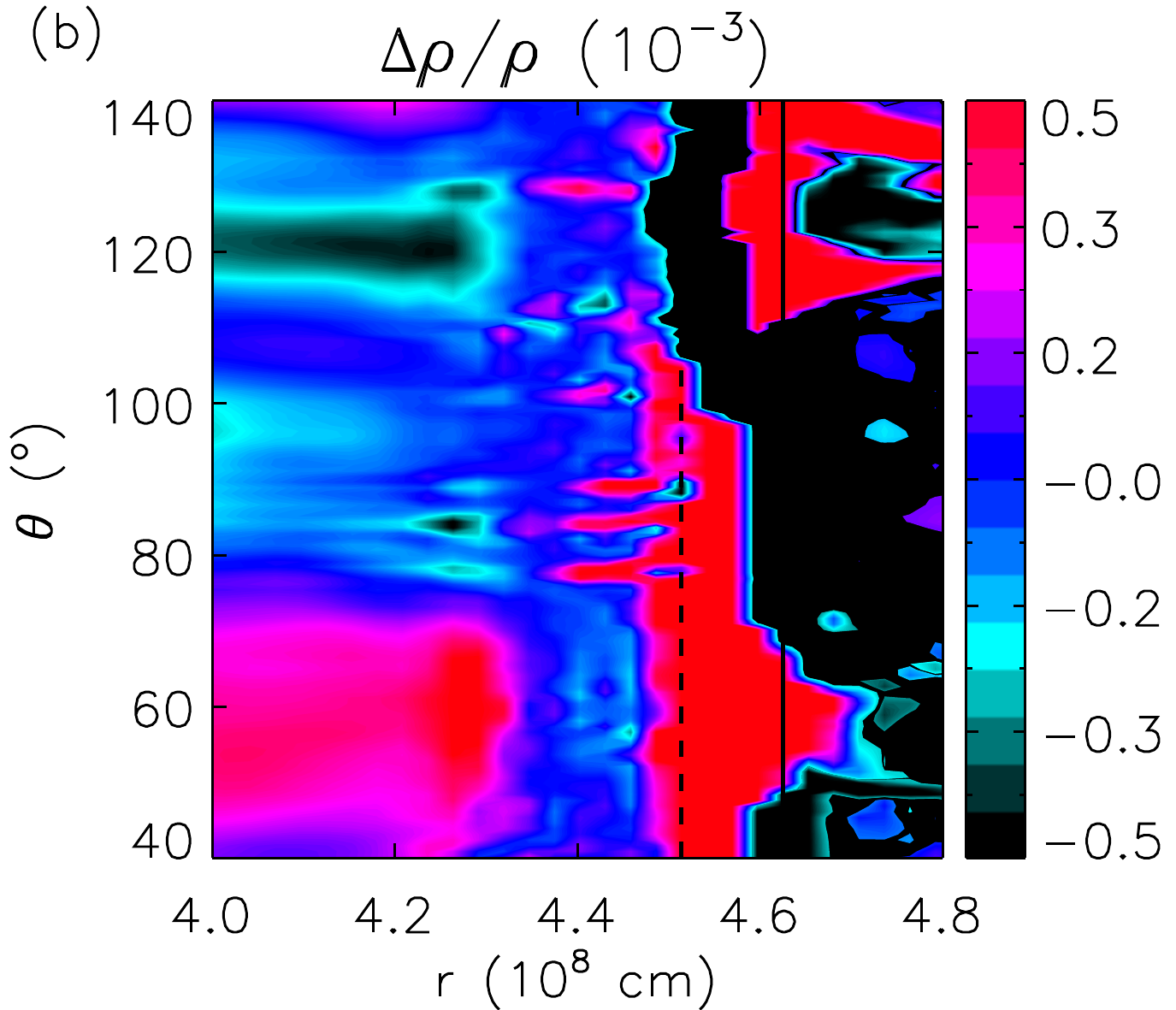}
\\
\includegraphics[width=0.47\hsize]{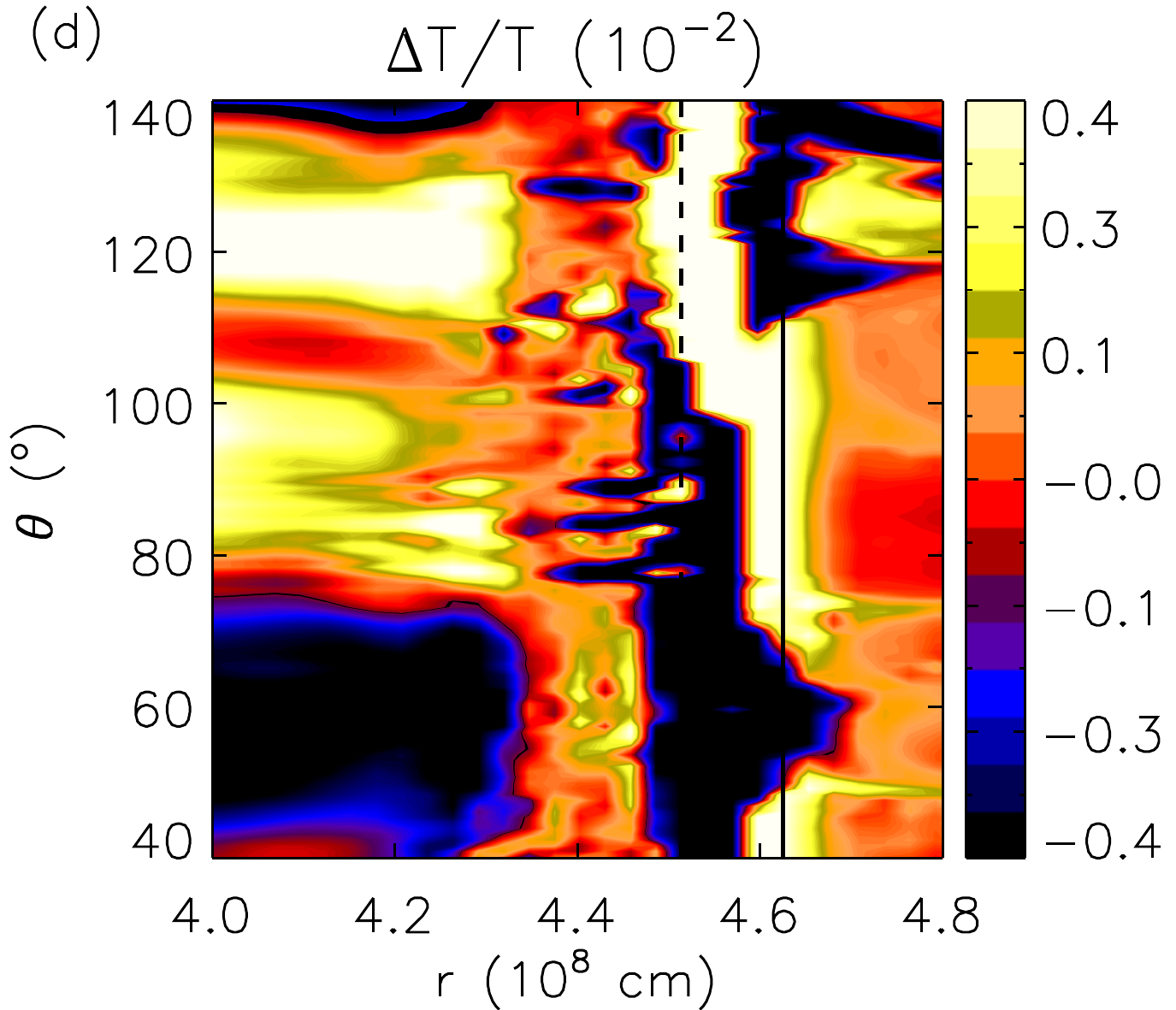}
\hfill
\includegraphics[width=0.47\hsize]{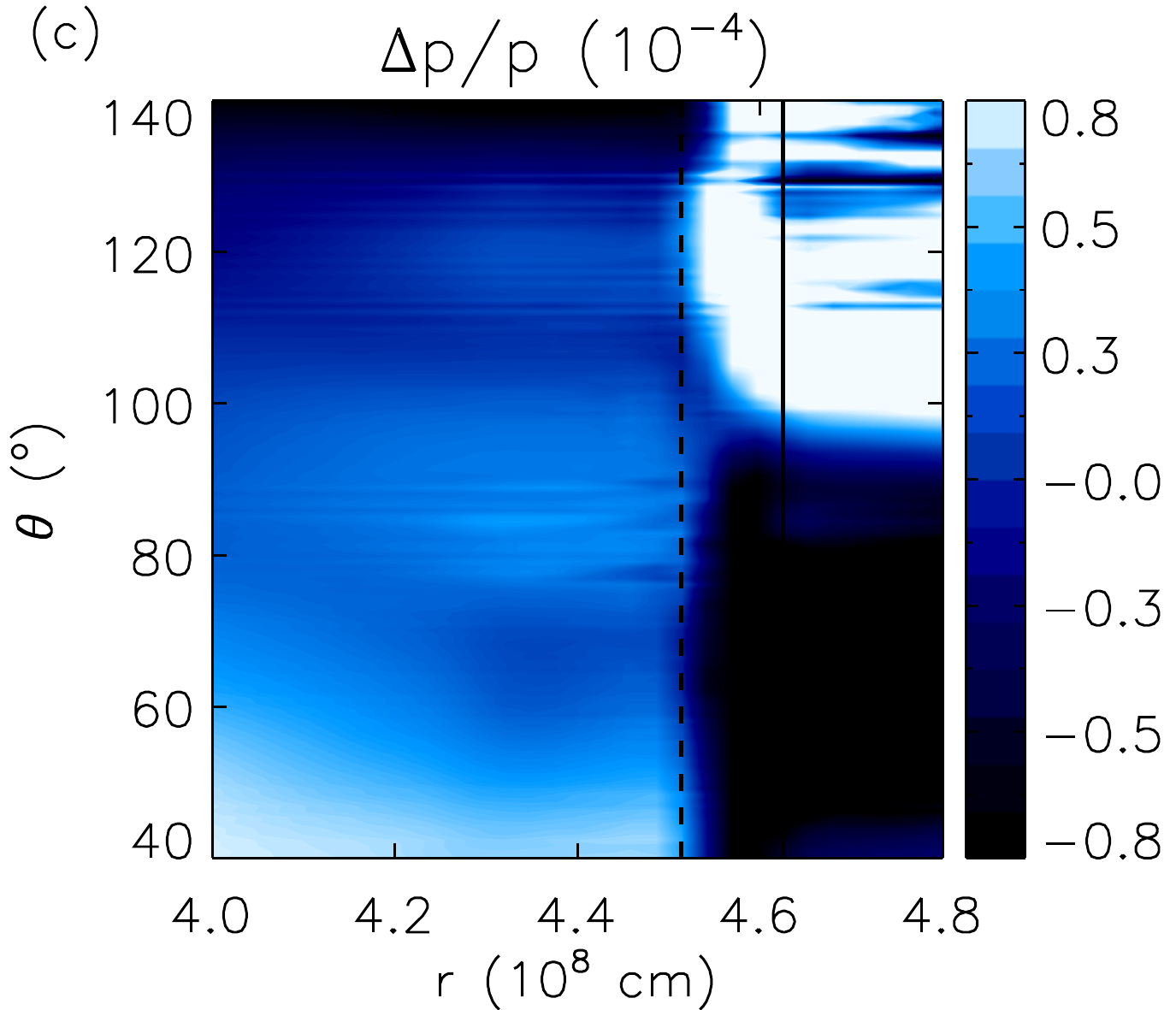}
\\
\includegraphics[width=0.47\hsize]{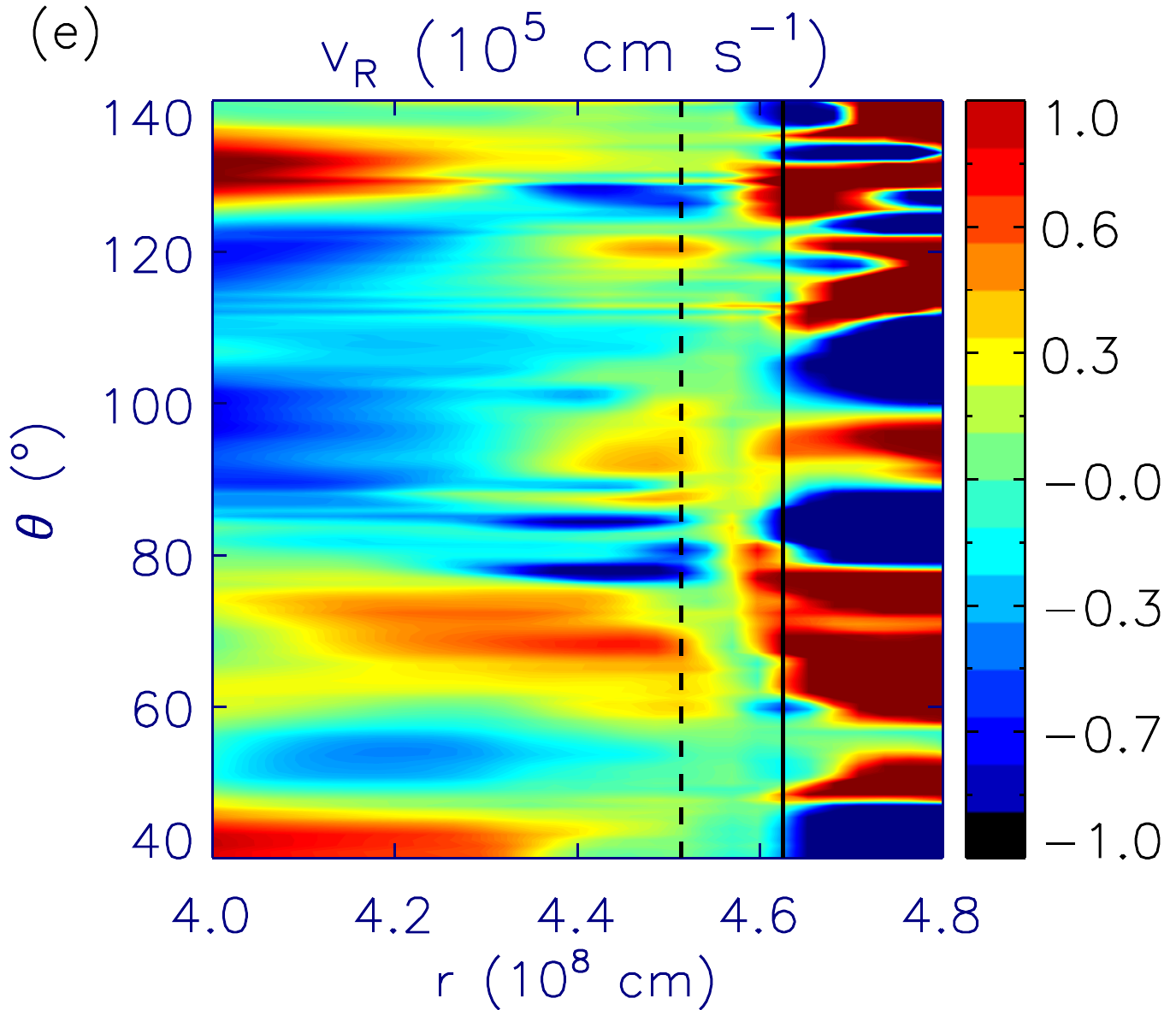}
\hfill
\includegraphics[width=0.47\hsize]{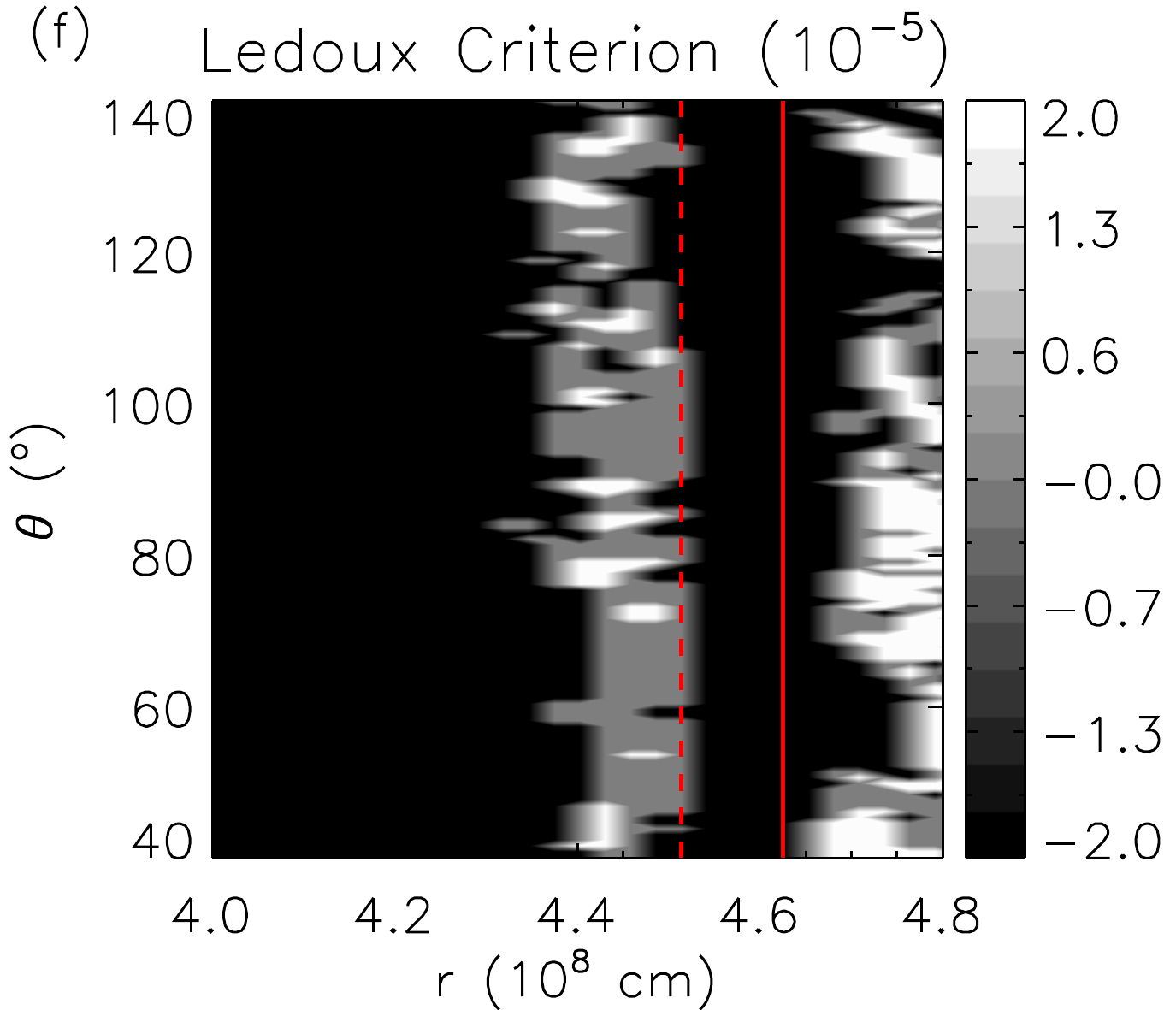}
\caption{Snapshots showing the base of the convection zone in the 2D
  model hefl.2d.3 at $t = 63430\,$s. The upper four panels give the
  relative difference between the local and the horizontally averaged
  value at a given radius of the mean molecular weight $\mu$, the
  density $\rho$, the temperature $T$, and the pressure $p$,
  respectively. The lower two panels display the radial velocity
  $v_R$, and the Ledoux criterion (a positive value implies that the
  flow is Ledoux unstable; see Eq.\,\ref{eq.ledoux}), respectively.
  The vertical solid line marks the bottom of the convection zone
  (location of $T_{max}$), while the dashed line gives the location
  from where RTFI mixing is launched. Note that only a part of the
  computational domain is shown. }
\label{fig.fingall}  
\end{figure*} 

\subsubsection{{{\color{red}}}Core carbon flash}
An analysis of the hydrodynamic properties of shell convection during
the core carbon flash, based on our 2D model cafl.2d, shows that the
convective flow is dominated by small circular vortices.  The maximum
angular averaged velocities inside the convection zone $|v| \sim
4\times 10^6 \cms $ exceed those predicted by MLT $v_{MLT} \sim 1.5
\times 10^5 \cms $ by about an order of magnitude.  This is an unusual
result, as in general, velocities inferred from 2D hydrodynamic
simulations of shell convection typically exceed MLT velocities by a
factor of five at most. The relative angular fluctuations of density
$\rho$ and temperature $T$ are of the order of $10^{-4}$ in the
convection zone, while those of the carbon mass fraction $X(^{12}$C)
are of the order of $10^{-3}$ (Fig.\,\ref{fig.heflcfloxfl}). Whether
turbulent entrainment is operative could not be decided, as the
simulated timescales are too short. However, we intend to address this
issue elsewhere.

\subsubsection{Oxygen shell burning}
The hydrodynamic properties of shell convection during oxygen shell
burning are discussed in detail in \citet{MeakinArnett2007}. The
findings of these authors are confirmed by our 2D model oxfl.2d very
well. The angular averaged amplitudes of the convective velocities
inferred from model oxfl.2d are $|v| \sim 1\times 10^{7}
\cms$. Turbulent entrainment has been detected in model oxfld.2d, too,
but we have not further analyzed this phenomenon; for more details see
\citet{MeakinArnett2007}. The relative angular fluctuations of density
$\rho$ and temperature $T$ are of the order of $10^{-4}$ in the
convection zone, while for example those of the oxygen mass fraction
$X(^{16}$O) are of the order of $10^{-3}$
(Fig.\,\ref{fig.heflcfloxfl}). The similarity of these values with
those in our models of the core helium and core carbon flash implies
that the fluctuations of thermodynamic quantities seem not to depend
on the specific type of shell convection.

\subsection{Mixing below the convection zone 
            in case of $\nabla_{\mu} < 0$}
\label{sect:mixingmuneg}

\begin{figure*} 
\includegraphics[width=0.99\hsize]{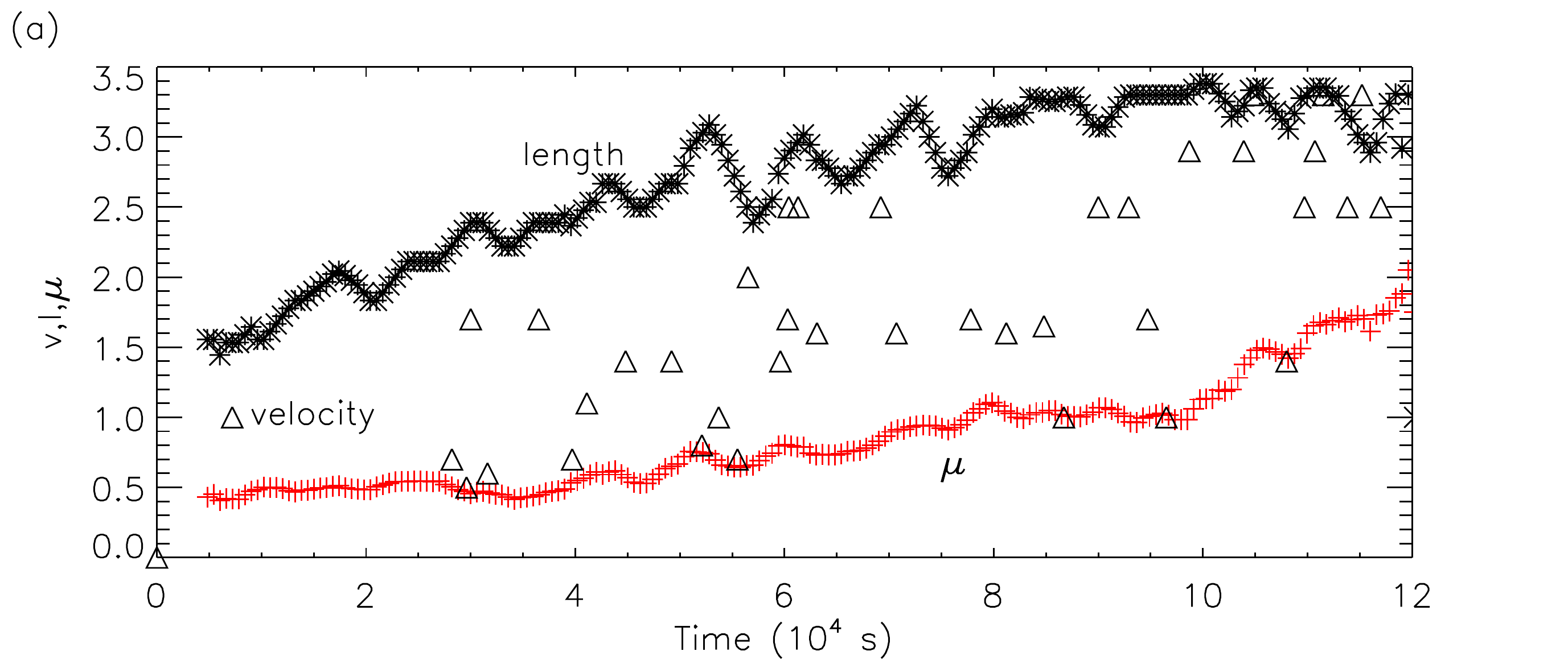}

\includegraphics[width=0.99\hsize]{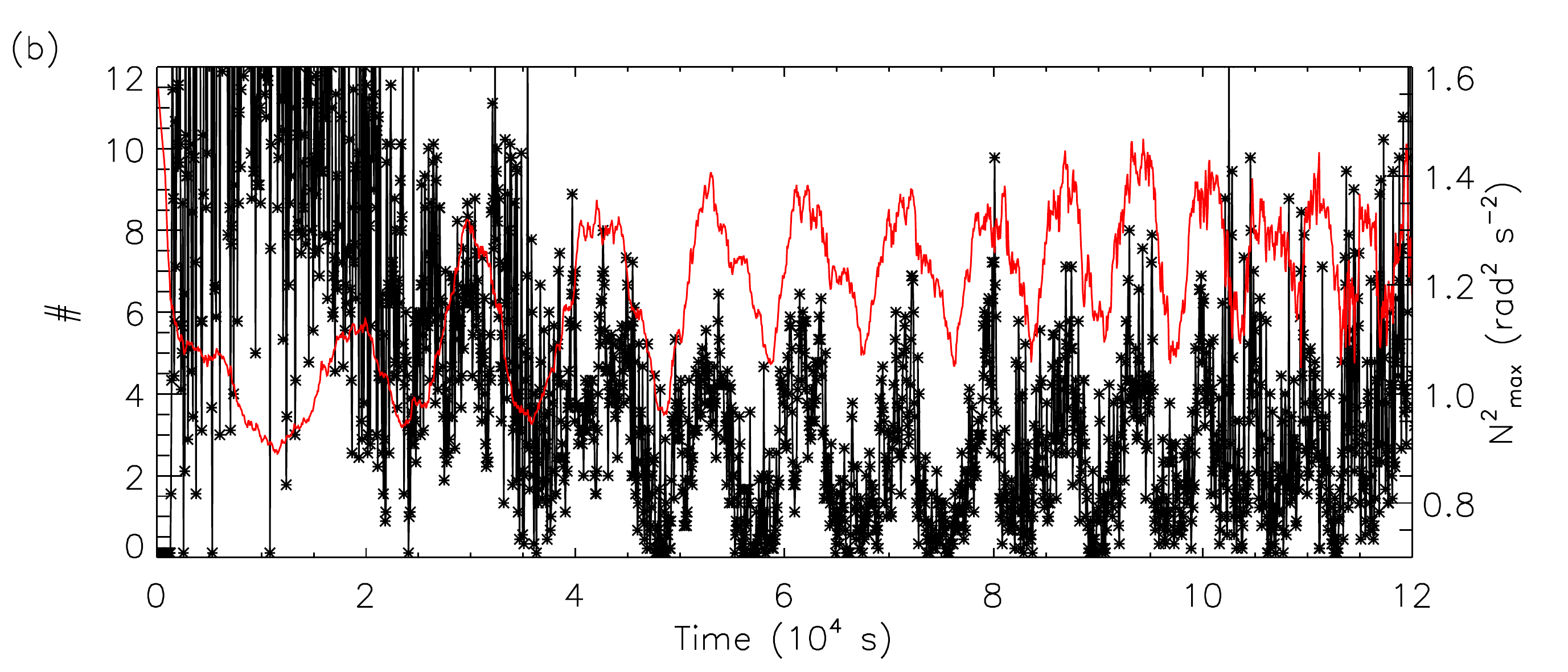}

\includegraphics[width=0.99\hsize]{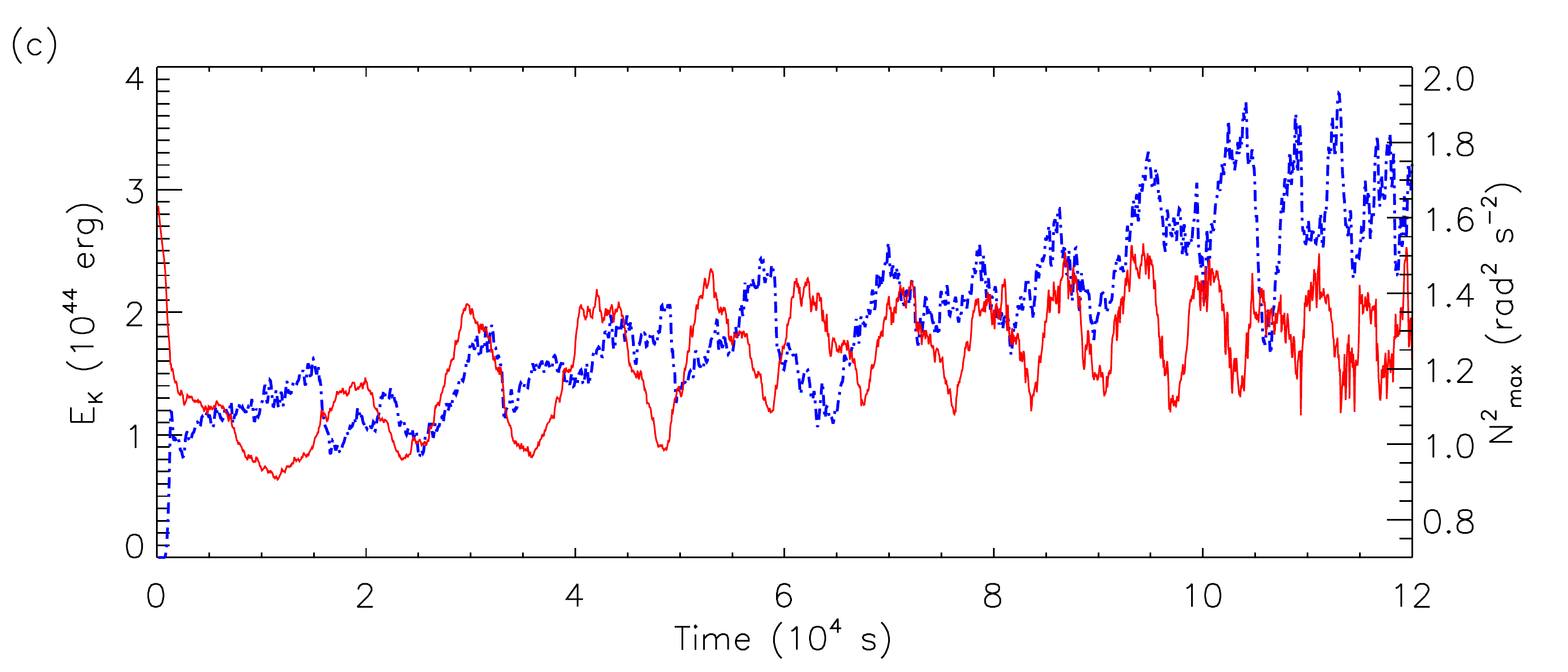}
\caption{Evolution of RTFI mixing in the 2D model hefl.2d.3. The upper
  panel shows the velocity in units of $10^5$\cms (triangles), the
  length of the fingers in units of $10^7\,$cm (stars), and the
  angle-averaged mean molecular weight $\mu - 4.066$ ($10^{-3}$) in the 
  region mixed by
  the RTFI (red crosses). The middle panel gives the number of RTFI
  fingers (black) together with the maximum value of the square of the
  Brunt-V\"ais\"al\"a frequency $N^2_{max}$ at the boundary of the
  convection zone (red). The lower panel displays the total kinetic
  energy $E_K$ within the convection zone (dash-dotted blue) together
  with $N^2_{max}$ (solid-red). }
\label{fig.fingtemp}
\end{figure*}

We discovered a new mixing process operating below shell convection
zones during degenerate, off-center helium and carbon burning (the
core helium and carbon flash, respectively).

The process is characterized by a steep negative mean molecular
weight gradient $\nabla_{\mu} < 0$ with $\nabla_{\mu} \equiv d \ln \mu
/ d \ln P$, where lighter gas with a higher mean molecular weight
resides above a denser gas of lower mean molecular weight. The
(negative) gradient results from nuclear burning in the convection
zone, and the assumption of instantaneous mixing used to calculate the
corresponding stellar models.

In the following, we present the thermodynamic characteristics of such
a mixing process during the core helium flash inferred from our 2D
model hefl.2d.3, which was simulated with the highest grid
resolution. Mixing is caused by overdense blobs enriched by material
from the convection zone (\eg carbon $^{12}$C) with higher mean
molecular weight $\mu$ literally shooting down from the base of the
convection zone towards the center of the star (\ie in the direction
of gravity). The blobs penetrate into denser layers of the core
thereby creating finger-like structures (Fig.\,\ref{fig.fingcarb}\,a,
\,b) which we also find in our 3D core helium flash model hefl.3d
(Fig.\,\ref{fig.fingcarb}\,c,\, d).  The properties of the mixing
found in our hydrodynamic core carbon flash model are
similar. Therefore, we discuss them only briefly at the end of this
section.

A detailed analysis of the flow in the layers just beneath the base of
the convection zone in model hefl.2d.3 revealed sinking dense blobs
enriched by material with higher mean molecular weight than that of
the ambient matter into which they penetrate.  Resulting finger-like
structures become visible immediately after convection starts in the
layers below the temperature maximum. The finger-like structures can
easily be spotted in maps showing relative angular fluctuations of the
mean molecular weight, density, and temperature, respectively
(Fig.\,\ref{fig.fingall}\,a - c).
\footnote{Note that in Figs.\,\ref{fig.fingcarb} and
  \ref{fig.fingall}, which show the elongated finger-like structures,
  the maps are displayed in spherical coordinates ($r$ and $\theta$ in
  cm and degree) using a rectangular projection.}
They grow continuously (Figs.\,\ref{fig.fingtemp}~a), and reach a
length of almost $2\times 10^7\,$cm within the first $20\,000\,$s of
evolution.  Their width remains almost constant at a value of roughly
1-2$\times 10^7\,$cm during the entire simulation.

\begin{figure} 
\includegraphics[width=0.91\hsize]{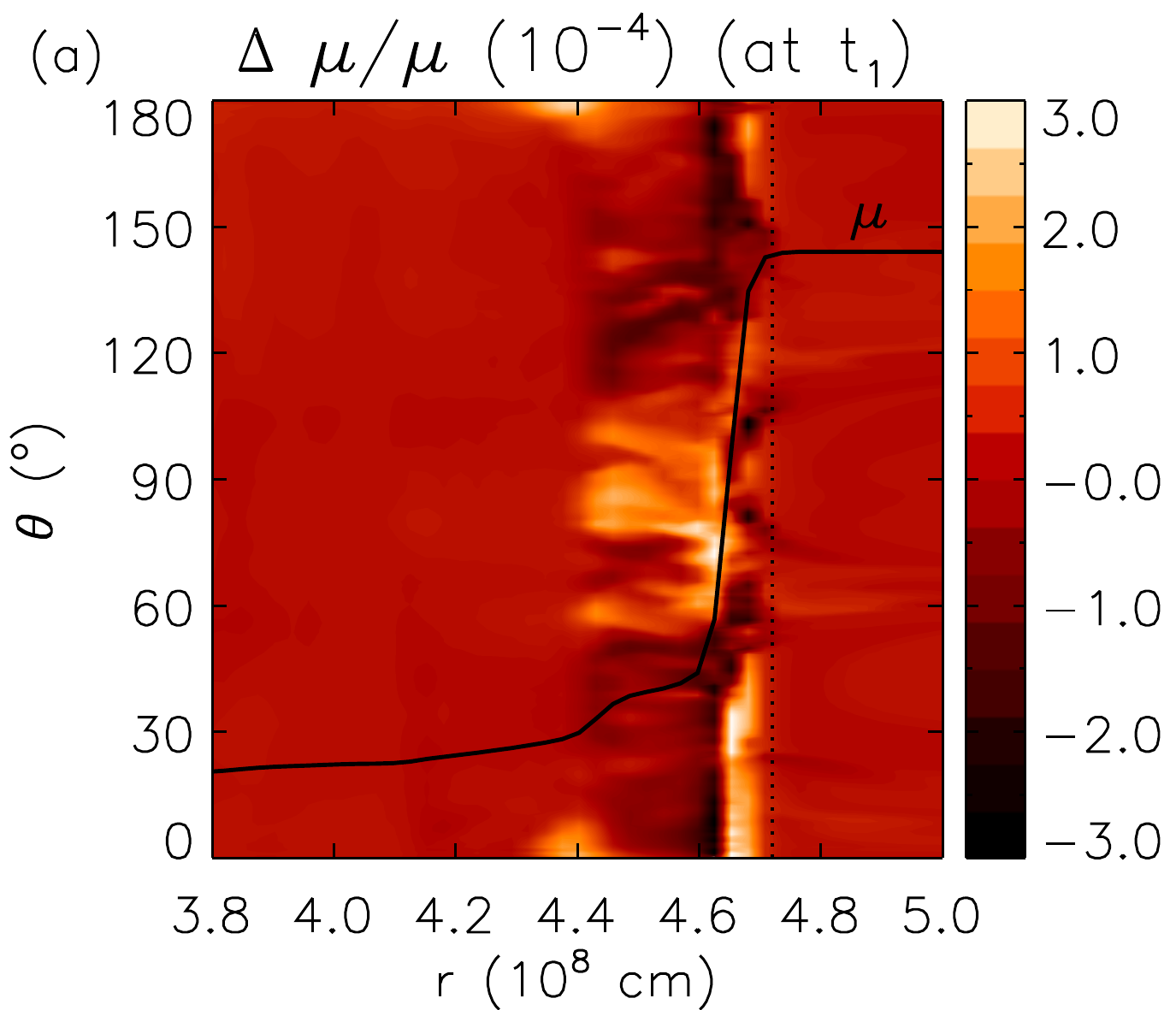}
\includegraphics[width=0.91\hsize]{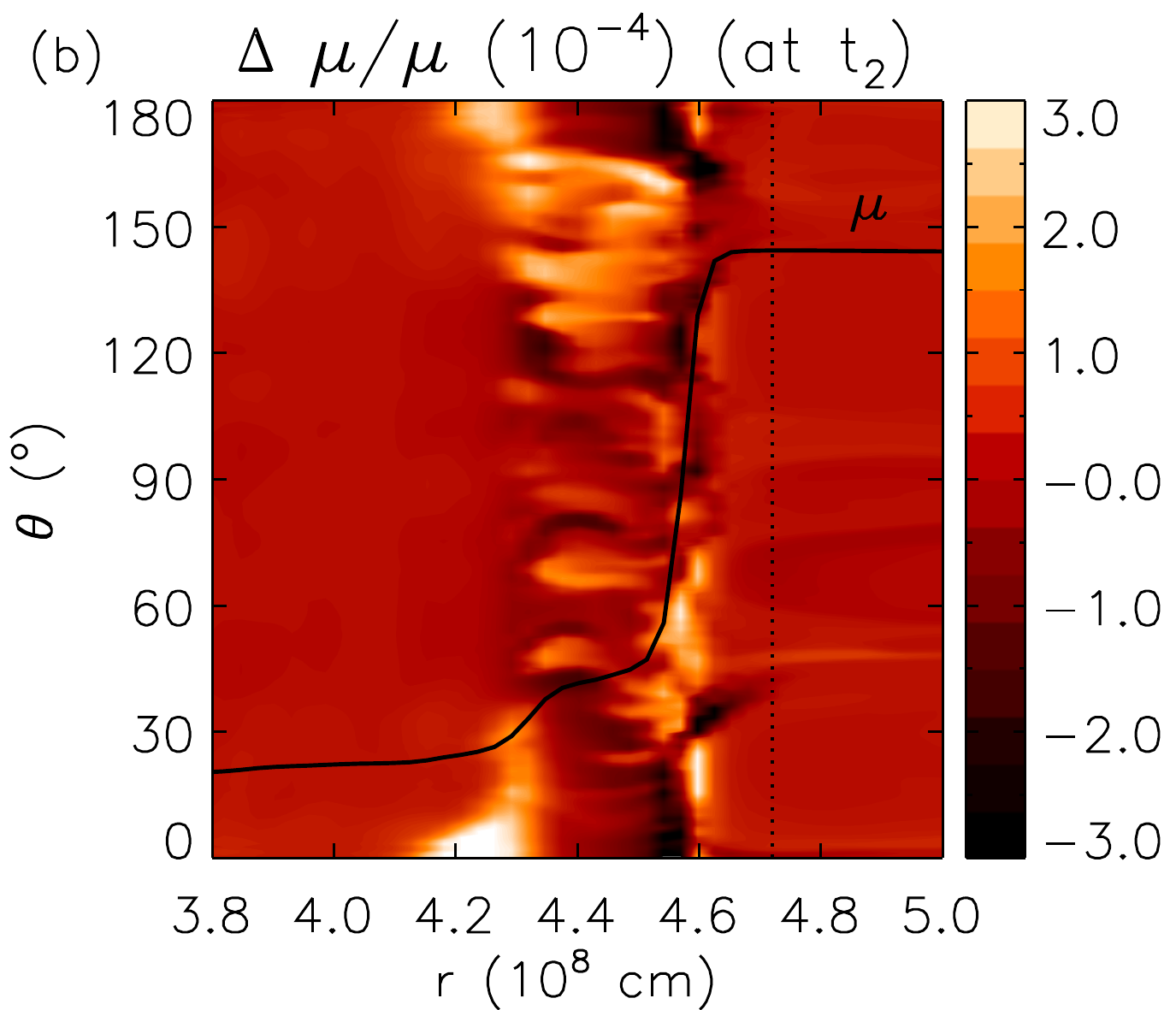}
\includegraphics[width=0.91\hsize]{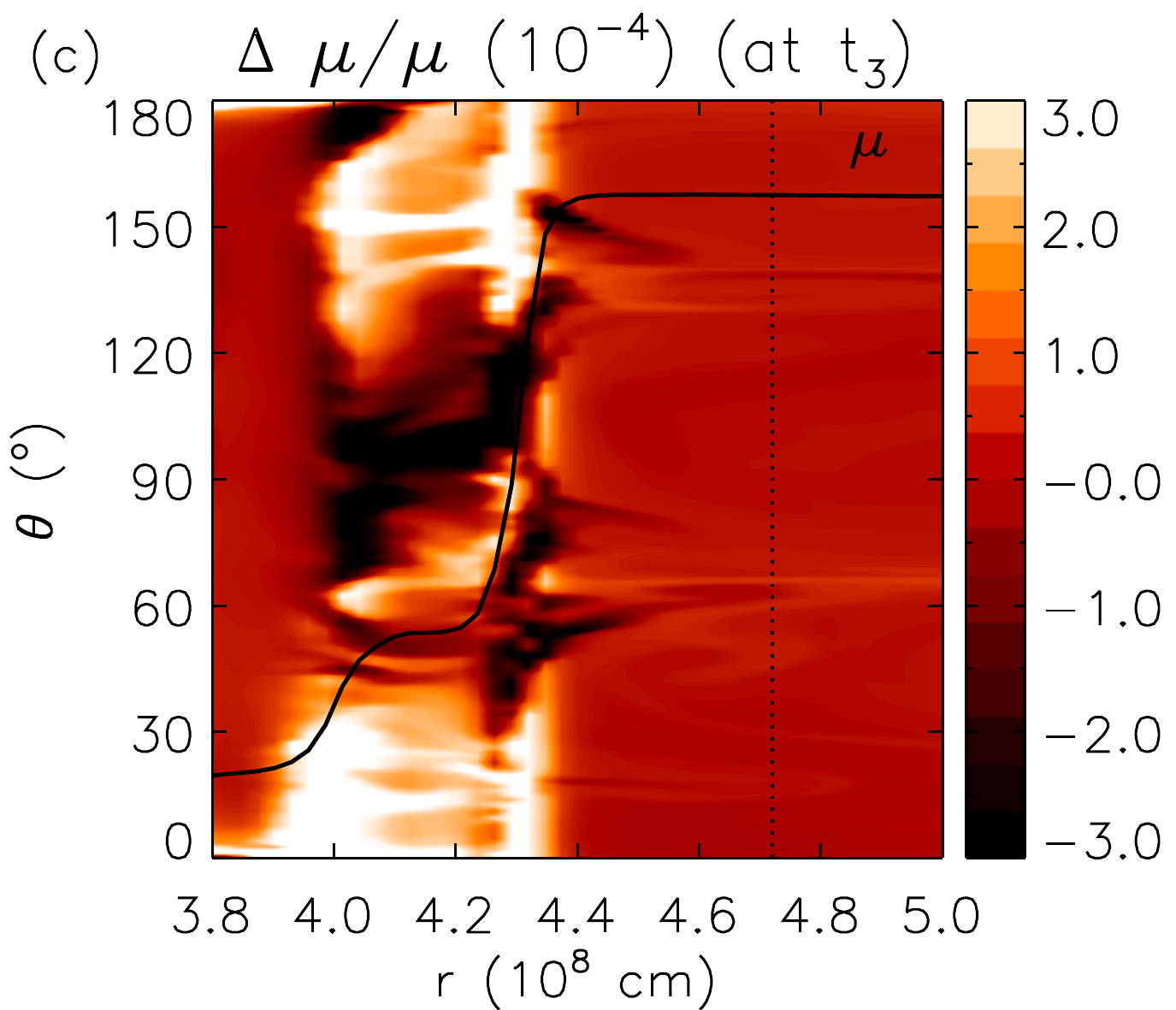}
\caption{Relative angular fluctuations of the mean molecular weight
  $\Delta \mu/\mu = (\mu - \langle \mu \rangle_{\theta}) \,/\, \langle
  \mu \rangle_{\theta}$ at the base of the convection zone in the 2D
  model hefl.2d.3 at $t_1 = 30\,000\,$s (a), $t_2 = 60\,000\,$s
  (b), and $t_3 = 120\,000\,$s (c), respectively. In each
  panel the black solid line gives the angular averaged radial
  distribution of the mean molecular weight $\mu$, and $\langle
  \rangle_{\theta}$ denotes the angular average at a given radius. Vertical
  dotted line marks the initial position of the convection boundary at t = 0~s 
  as determined by Schwarzschild criterion.}
\label{fig.fingvistmp}  
\end{figure}

Although, the finger-like structures created by heavy sinking blobs of
gas at earlier times eventually diminish
\footnote{The typical lifetime of a sinking heavy blob is about 350\,s
  when our simulations have progressed for about 60\,000\,s.},
they are being constantly replaced by new heavy blobs again leaving
behind trails of gas enriched by matter of larger $\mu$. Thus, the
layers into which the fingers penetrate are enriched by material of
higher $\mu$.  Consequently, the initial value of $\mu = 4.06633$
rises by $0.05\%$ below the convective shell until the end of the
simulation (Figs.\,\ref{fig.fingtemp}a and \ref{fig.fingvistmp}) at a
steadily increasing rate
\footnote{At the peak of the flash $\mu$ rises at a rate of $2 \times
  10^{-8}\,$s$^{-1}$ within the convection zone, which is only twice
  faster than the rate of $\mu$ enrichment due to RTFI mixing below
  the convective shell.}.

The sink velocities of the overdense blobs are initially roughly
$1\times 10^5 \cms$ reaching approximately $3\times 10^5 \cms$ at the
end of the simulation at $t = 33.3\,$hrs. The blobs are always cooler
and denser than the ambient matter by roughly $0.4\%$ and $0.05\%$,
respectively. Thus, the observed finger-like structures cannot be
caused by the double diffusive instability which leads to
"salt-fingers", as these are hotter than the surrounding matter
leading to heat diffusion that is fast compared to diffusion of
composition \citep{Schmitt1994}.

Using the size of relative density fluctuations $\Delta \rho / \rho$
within the region where RTFI mixing takes place we can derive an
analytic estimate for the velocity of the dense blobs causing the
finger-like structures, if we assume that the blobs arise from a
Rayleigh Taylor instability \citep{CoxGiuli2008}. With $\Delta \rho /
\rho \sim 5\times 10^{-4}$, inferred from our simulation
(Fig.\,\ref{fig.fingall}\,a), we find
\begin{equation}
  v^2 \sim g \Lambda \frac{\Delta \rho}{\rho} \sim 7\times 10^5 \cms  
\label{eq:deltarho1}
\end{equation}
where $g \sim 10^8 \cmss$ is the gravitational acceleration at the
base of the convection zone, and $\Lambda \sim 10^7 \cm$ the
approximate length of the fingers (Fig.\,\ref{fig.fingtemp}).

The above estimate agrees well with the results of an analysis of our
simulation, where the gas inside the fingers sinks at a bit smaller
velocities ranging from $\sim 1 \times 10^{5}$\cms up to $3.5 \times
10^{5}$\cms (Fig.\,\ref{fig.fingall} and \ref{fig.fingtemp}).  It
confirms that the ``buoyancy work'' ($\sim v^2$) is consistent with
the acceleration of gas seen in the simulation, and suggests that the
mixing is the result of the Rayleigh-Taylor instability. Hence, we
propose to call it Rayleigh-Taylor finger instability mixing (RTFI).
The velocities observed inside the finger-like structures
(Fig.\,\ref{fig.fingall}\,e) are smaller than those predicted by
Eq.\,(\ref{eq:deltarho1}), likely because of the friction caused by
our numerical scheme between the ambient medium and the sinking
overdense blob.

\begin{figure}
\includegraphics[width=0.93\hsize]{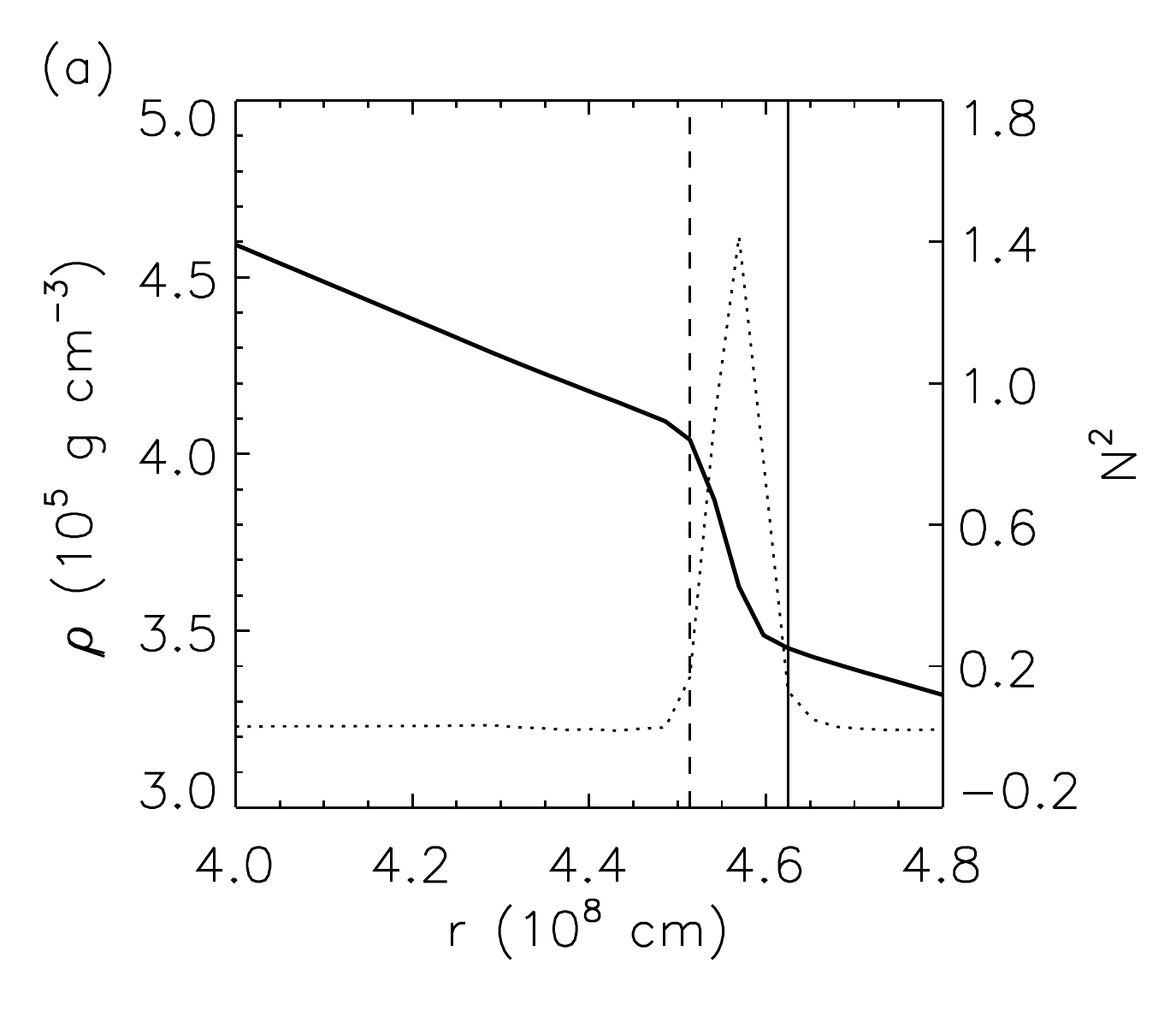}
\vspace{-0.2cm}
\includegraphics[width=0.93\hsize]{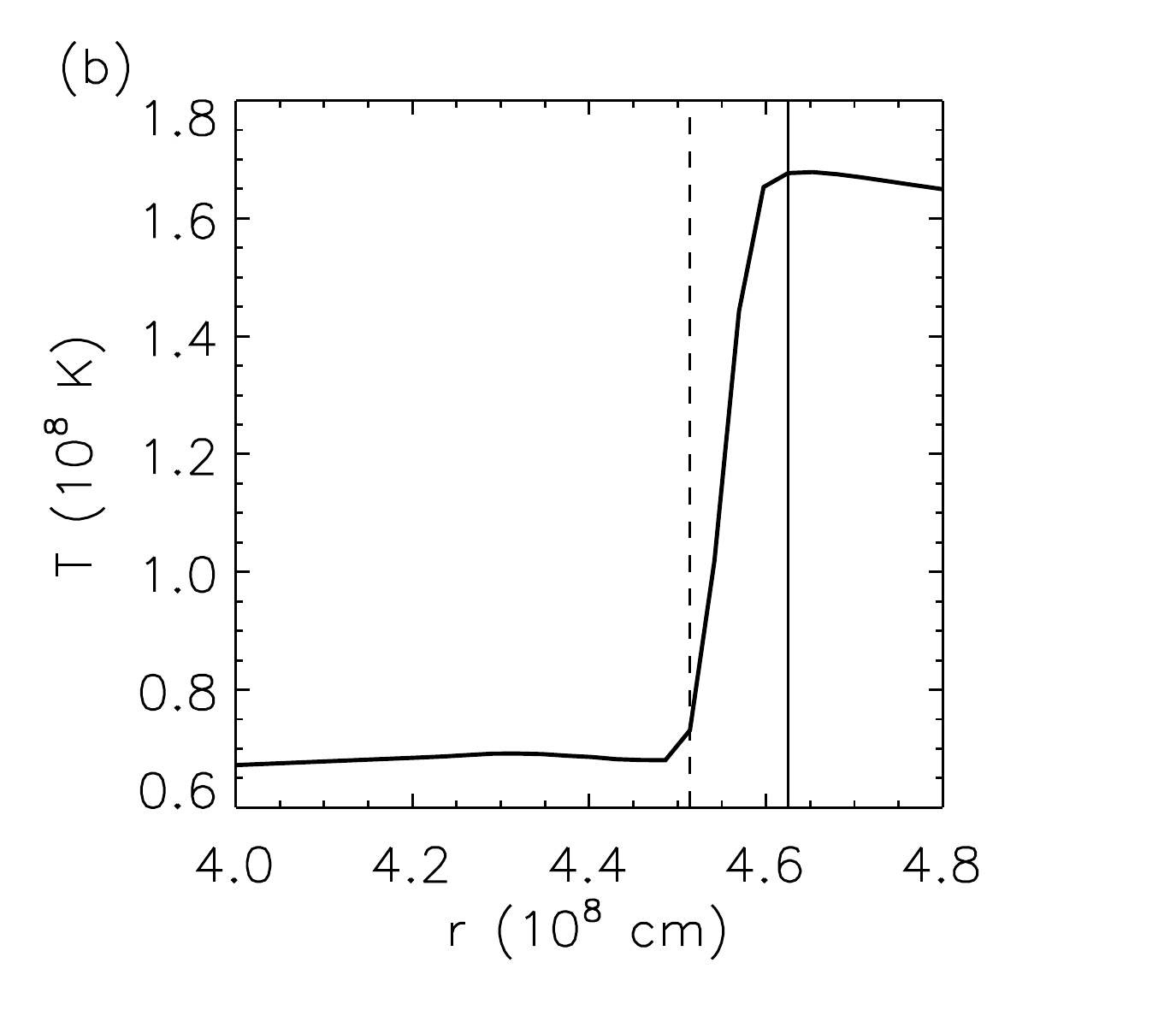}
\vspace{-0.2cm}
\includegraphics[width=0.93\hsize]{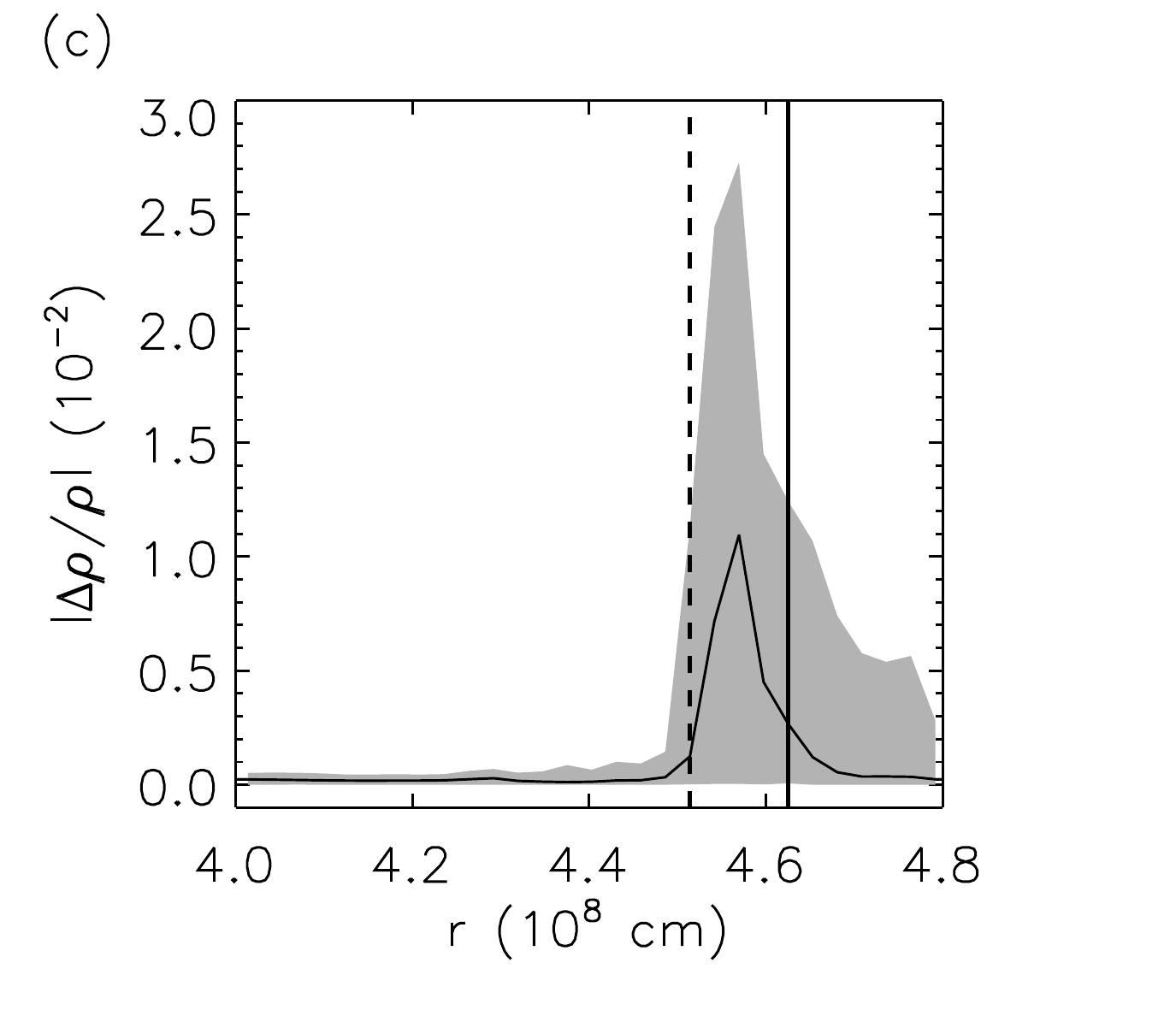}
\caption{Snapshots from the 2D model hefl.2d.3 at $t = 63430\,$s
  showing (a) the angular averaged density (solid) and the square of
  the Brunt-V\"ais\"al\"a frequency $N^2$ (dotted), (b) the angular
  averaged temperature, and (c) the modulus of the relative density
  fluctuation $|(\rho-\langle \rho \rangle_\theta)/\langle \rho
  \rangle_\theta|$ (including its angular variation; shaded) as a
  function of radius near the bottom of the convection zone.  The
  vertical solid line marks the bottom of the convection zone
  (location of $T_{max}$), while the dashed line gives the location
  from where RTFI mixing is launched. }
\label{fig.fingall1}
\end{figure}

Note that the initial density fluctuations can result from wrinkling
of the contact discontinuity at the boundary of the convection zone
due to the convective flow, turbulence, g-modes or p-modes.

\citet{Eggleton2006} also found Rayleigh-Taylor instabilities driven
by a molecular weight inversion caused by the
$^{3}$He\,$(^{3}$He,2p)\,$^{4}$He reaction in hydrodynamic simulations
of the hydrogen-rich layers residing above the hydrogen burning shell
in low-mass red giants. The corresponding mixing is qualitatively
similar to RTFI mixing. The mixing velocities observed by
\citet{Eggleton2006} agree very well with those theoretically
estimated for a RT instability according to the formula $v^2 \sim g
H_p\, \Delta\mu / \mu$, where equal relative density and composition
fluctuations, \ie $\Delta \rho / \rho = \Delta\mu / \mu$, were
assumed. However, any density fluctuation will necessarily lead to
composition and temperature fluctuations $\Delta T$ as well. Moreover,
the formula of \citet{Eggleton2006} holds only for an ideal gas, and
only if the blobs are in pressure equilibrium with the ambient medium
and have the same temperature as the surrounding matter.

The finger-like structures resulting from RTFI mixing in our
simulations are in pressure equilibrium with the surrounding gas, as
they move with subsonic velocities (the speed of sound is $c_s \sim
10^8 \cms$ in the corresponding layers). However, they are always
cooler than the ambient medium, \ie $T > T_1 \equiv T + \Delta T$,
where $T$ is the temperature of the ambient medium. Allowing also for
fluctuations of the mean molecular weight $\Delta \mu$, pressure
equilibrium ($p \equiv p_1)$ implies for an ideal gas equation of
state
\begin{equation}
 \frac{\rho T}{\mu} = \frac{(\rho + \Delta \rho) (T + \Delta T) }{ 
                      \mu + \Delta \mu } \, . 
\label{eq:deltarho-aux1}
\end{equation}
Solving for $\Delta \rho / \rho$ this relation should provide an 
improved estimate for the relative density variation in
Eq.\,(\ref{eq:deltarho1}):
\begin{equation}
 \frac{\Delta \rho}{\rho} = 
 \frac{T\, (\mu+\Delta\mu)}{\mu\, (T+\Delta T)} - 1 =
 \frac{1+\Delta\mu/\mu}{1+\Delta T/T} - 1 \, .
\label{eq:deltarho2}
\end{equation}

However, in our models RTFI mixing operates in a partially
electron-degenerate core, \ie the assumption of an ideal gas equation
of state is not justified as it was the case in \citet{Eggleton2006}.
Using the formula of \citet{Eggleton2006} the velocity of the sinking
overdense blobs of RTFI mixing can be estimated to be v$~\sim 5 \times
10^5 \cms$ ($\Delta \mu/\mu \sim 3\times10^{-4}$), while
Eqs.\,\ref{eq:deltarho1} and \ref{eq:deltarho2} lead to the
considerably larger value $v \sim 2\times 10^6 \cms$ ($\Delta \mu/\mu
\sim 3\times10^{-4}$, $\Delta T/T \sim -4\times 10^{-3}$), because the
density contrast inferred from Eq.\,\ref{eq:deltarho2}
$\Delta\rho/\rho = 4\times 10^{-3}$ is roughly an order of magnitude
larger than that observed in the simulations.

We point out here once more that although the finger-like structures
resemble overshooting, they actually originate from density
fluctuations just beneath the lower boundary of the convection zone
(Fig.\,\ref{fig.fingall1} a) where the gas is already cold compared to
that in the convection zone (Fig.\,\ref{fig.fingall1} b), but still
partially enriched by matter from the adjacent convection zone. The
fingers are also neither directly connected with sinking plumes inside
the convection zone nor with large density fluctuations seen at the
lower convection zone boundary (Fig.\,\ref{fig.fingall1} c).

\begin{figure*} 
\includegraphics[width=0.47\hsize]{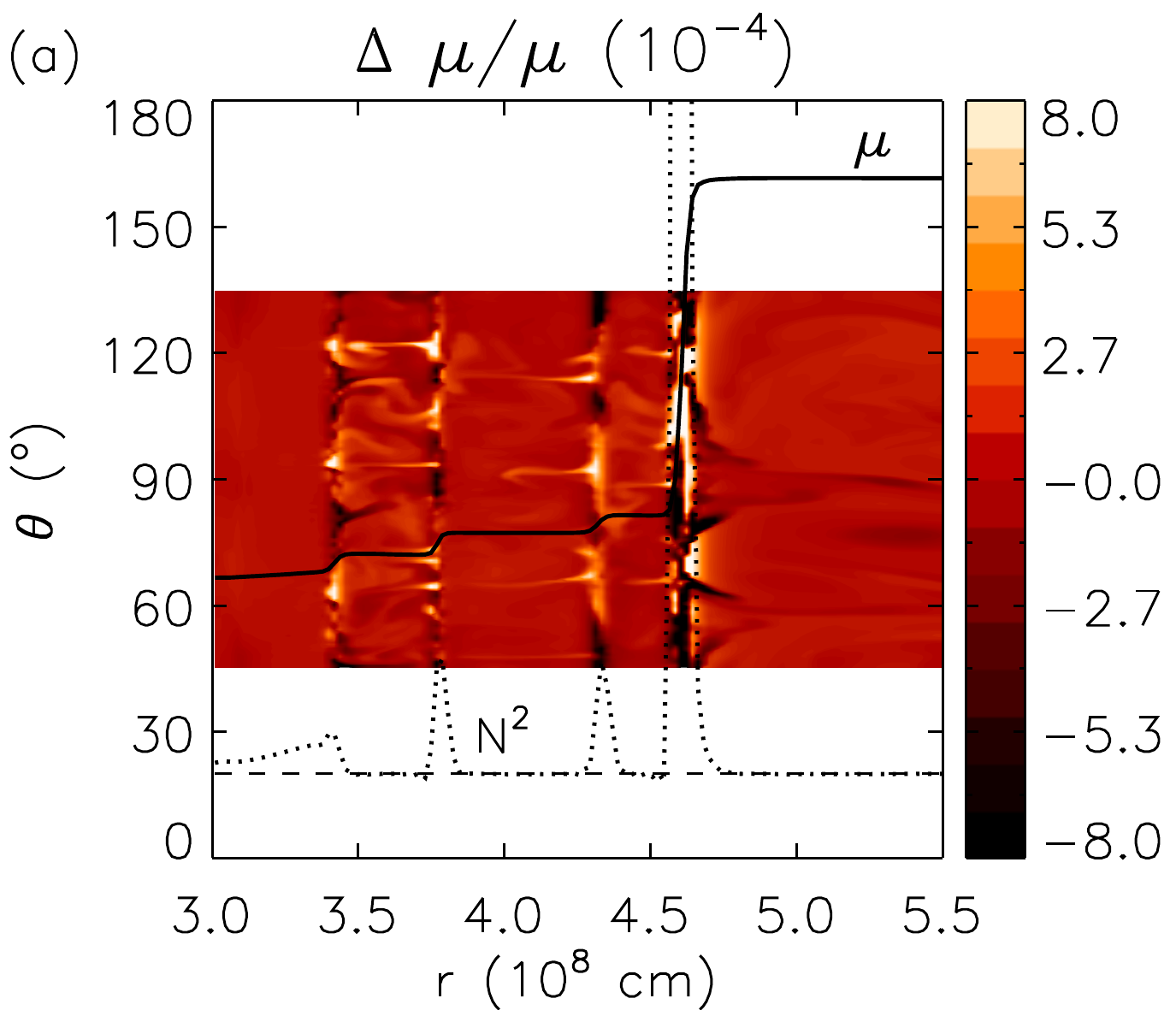}
\hfill
\includegraphics[width=0.47\hsize]{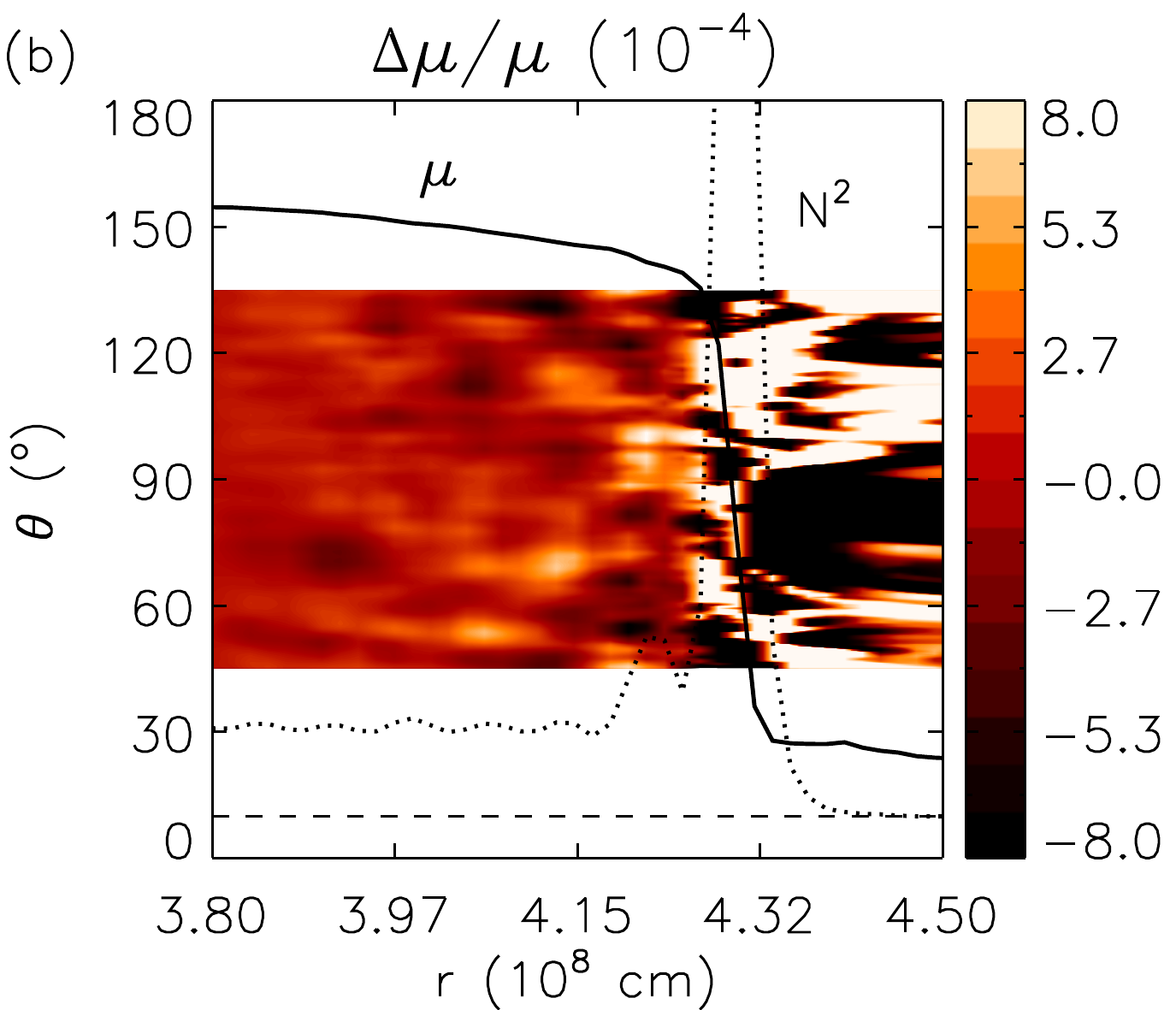}
\caption{Relative angular fluctuations of the mean molecular weight
  $\Delta \mu/\mu = (\mu - \langle \mu \rangle_{\theta}) \,/\, \langle
  \mu \rangle_{\theta}$ at the base of the convection zone in the 2D
  model cafl.2d at $t = 10230\,$s (a), and in the 2D model oxfl.2d
  at $t = 438\,$s (b), respectively. In both panels the dotted
  line shows the angular averaged distribution of the square of the
  Brunt-V\"ais\"al\"a frequency $N^2$, and the solid line the angular
  averaged mean molecular weight $\mu$. The horizontal dashed line
  corresponds to $N^2 = 0$, and $\langle \rangle_{\theta}$ denotes the
  angular average at a given radius.}
\label{fig.cofing}
\end{figure*}

There exists a correlation
\footnote{The correlation is not very clear during the first $2\times
  10^4\,$s, as it is difficult to determine the number of fingers
  during this early epoch. To identify a finger one searches for
  overdense regions below the base of the convective shell that
  correspond to sinking blobs. Focusing on fluctuations in $\mu$ is
  misleading, as these often exist only as traces behind sinking heavy
  blobs.}
between the maximum value of the square of the Brunt-V\"ais\"al\"a
frequency $N^2_{max}$ and the number of new fingers (heavy blobs)
appearing at a given time (Fig.\,\ref{fig.fingtemp}\,b).  $N^2_{max}$
varies between $1 \radss$ and $1.5 \radss$ for the 2D model
hefl.2d.3. This can be understood, as the fingers are caused by
overdense blobs of gas and the Brunt-V\"ais\"al\"a frequency can be
related to the amplitude of the density oscillations according to
\begin{equation}
  N^{2} = \frac{g \delta}{H_P} \left( \nabla_{ad} - \nabla_{sur} + 
          \frac{\phi}{\delta} \nabla_{\mu} \right) 
          \quad \Longrightarrow \quad 
          \frac{\Delta \rho}{\rho} = \frac{N^2}{g} \Delta r 
\label{eq.brunt1}
\end{equation}
Another well known interpretation of $N^2$ relates it to the frequency
of g-modes or internal gravity waves:
\begin{equation}
  N^{2} = \frac{g \delta}{H_P} \left( \nabla_{ad} - \nabla_{sur} + 
          \frac{\phi}{\delta} \nabla_{\mu} \right) 
          \quad \Longrightarrow \quad 
          \Delta r = \Delta r_0 e^{iNt} \, ,
\label{eq.brunt}
\end{equation}

\noindent
where $\nabla_{sur} = d \ln T / d \ln P$ is the logarithmic
temperature gradient in the ambient surrounding, $\nabla_{ad} = (d \ln
T / d \ln P)_{ad}$ gives the temperature change for an adiabatic
displacement, and $\nabla_{\mu} = d \ln \mu /d \ln P$ the
corresponding logarithmic composition gradient. Furthermore, we have
$\delta = -\partial \ln \rho \,/\, \partial \ln T$ and $\phi =
\partial \ln \rho \,/\, \partial \ln \mu$, where the partial
derivatives are evaluated at constant values of $(P, \mu)$ and $(P,
T)$, respectively. Finally, $H_p$ is the pressure scale height.

The correlation tells us that larger density fluctuations (larger
values of $N^2$) result in the formation of denser and therefore
heavier blobs of gas at the bottom edge of the convection zone, which
sink more easily down into the core.  Such blobs do not have to be
extremely heavy (dense) initially. It is sufficient when their density
remains higher than that of the ambient medium on length scales
comparable with the size of the observed finger-like structures.  The
dense blobs sink down from the bottom edge of the convection zone to
layers, which are about $\sim 20\%$ denser than those from where the
blobs are created (Fig.\,\ref{fig.fingall1}).  This is possible as the
sinking blobs are compressed, because they remain in pressure
equilibrium with the ambient medium. Thus, the situation is exactly
opposite to that encountered in the case of convection, where hot
blobs of relatively dense gas rise from the bottom of the convection
zone over large distances because they expand, although their initial
density is much higher than that of the layers where they eventually
end up.

No or only a weak correlation can be recognized between the flow in
the convection zone and the occurrence of sinking dense blobs (causing
the finger-like structures) when comparing the temporal evolution of
the total kinetic energy of the convective flow and the number of
fingers appearing at a given time (up to six) or the maximum of the
Brunt-V\"ais\"al\"a frequency squared (Fig.\,\ref{fig.fingtemp}\,c).
However, there exists a correlation between the positions of the hot
spots at the base of the convection zone and the locations where the
fingers appear.  Fingers occur more frequently at the positions of the
temperature maxima (by $\sim 40\%$) as compared to any other place
beneath the convection zone, the temperature maxima being typically
connected with convective upflows.

We have no explanation for the observed periodic behavior of
$N^2_{max}$, and for the occurrence period of the finger-like
structures (Fig.\,\ref{fig.fingtemp}\,b), which is roughly $10^4\,$s
(or $\sim 20$ convective turnover timescales) at $t \sim 60\,000\,$s
and decreases at later times.  The periodicity seems to be
uncorrelated with the convection as the total kinetic energy of the
convective flow and $N^2_{max}$ are not clearly correlated
(Fig.\,\ref{fig.fingtemp}\,c).

We have also found RTFI mixing in multidimensional models of shell
convection during the core carbon flash, which suggested to us that a
crucial ingredient for the operation of the instability is the
presence of a composition gradient, not necessarily coinciding with
the bottom of convection zone. During the core carbon flash mixing
leads to the formation of several layers enriched by matter of a
larger mean molecular weight separated by interfaces with high values
of $N^2$ (Fig.\,\ref{fig.cofing},~a) indicating large density
fluctuations (Eq.\,\ref{eq.brunt1}).  The bottom two layers are
completely detached from the convection zone whose lower edge is
located at $r \sim 4.6\times 10^8\,$cm. The radial distribution of
$\mu$ across these layers shows a staircase-like behavior, which is
similar to that observed in experiments of thermohaline mixing, \ie a
thermohaline staircase \citep{Krishnamurti2003}.

\subsection{Mixing below the convection zone 
            in case of $\nabla_{\mu} > 0$ }
\label{sect:mixingmupos}

In order to test our preliminary conclusion based on 2D shell
convection models of the core helium flash and the core carbon flash,
which both have a negative composition gradients at the base of the
convection zone, we have simulated in addition a 2D hydrodynamic model
of shell convection during the oxygen shell burning in a massive star.
The corresponding stellar model is characterized by a layer having a
positive mean molecular weight gradient $\nabla_{\mu} > 0$ at the base
of the convection zone. For this situation we do not find the
finger-like structures of RTFI mixing beneath the base of the
convection zone ($4.15 \times 10^8\,\mbox{cm} < r < 4.32 \times
10^8\,$cm) (Fig.\,\ref{fig.cofing}\,b).

\section{Stability Versus Local Perturbations}
\label{sect:locstab}

For assessing the dynamic stability of a stellar layer we consider a
fluid element which is radially displaced by an amount $\Delta r$ from
its initial position \citep{KipWeigert1990}. The density of the fluid
element will differ at its new position from that of its surrounding
by
\begin{equation}
 D\rho = \left[ \left( \frac{d\rho}{dr} \right)_{ele} - \, 
         \left( \frac{d\rho}{dr} \right)_{sur} \right]\, \Delta r
\end{equation}

\noindent
where $(d\rho/dr)_{ele}$ is the density change of the fluid element
when it rises a distance $dr$, and $(d\rho/dr)_{sur}$ is the density
gradient of the surrounding matter.

According to \eg \citet{KipWeigert1990} a stellar layer is unstable
against convection, if a fluid element that is initially lighter than
its surrounding will continue to rise when it is displaced against the
direction of gravity (\ie $\Delta r > 0$). Hence convection sets in
when the following criterion is fulfilled:
\begin{equation}
  \left( \frac{d\rho}{\rho} \right)_{ad} - 
  \left( \frac{d\rho}{\rho} \right)_{sur}    < 0 \, ,
\label{eq.densled}
\end{equation}

\noindent
where the differentials refer to changes in radial direction. If 
it is assumed that the fluid element moves sufficiently fast, it 
moves adiabatically, and hence $(d\rho/dr)_{ele} =
(d\rho/dr)_{ad}$.  Note that the same criterion also holds for a fluid
element that is initially heavier than its surrounding, and hence
sinks downward (\ie $\Delta r < 0$) in the direction of gravity.  Note
further that such an over-dense fluid element will sink downward as
long as its density remains higher than that of the surrounding
matter.

Allowing also for (radial) composition gradients within the stellar
layers Eq.\,\ref{eq.densled} leads to the well known Ledoux instability
criterion (see \eg, \citet{KipWeigert1990})
\begin{equation}
  \nabla_{sur} - \nabla_{ad} - \frac{\phi}{\delta}\nabla_{\mu} > 0 \, ,
\label{eq.ledoux}
\end{equation}

\noindent
where $\nabla_{sur}$, $\nabla_{ad}$, $\nabla_{\mu}$, and
$\phi,~\delta$ are defined above (see paragraph after
Eq.\,\ref{eq.brunt}).

The Ledoux criterion determines dynamically unstable regions in a
star.  Hence, if we could identify layers in our hydrodynamic models
beneath the base of the convection zone where the inequality
Eq.\,\ref{eq.ledoux} holds, we could prove that the over-dense blobs
which start sinking downward from there and causing RTFI mixing, are
the result of a Ledoux unstable stratification.  Indeed we find that
the left term of Eq.\,\ref{eq.ledoux} \ie
$\nabla_{sur}-\nabla_{ad}-\frac{\phi}{\delta}\nabla_{\mu}$ is positive
in our 2D core helium flash model in most grid zones where we observe
RTFI mixing (see white regions in Fig.\,\ref{fig.fingall}\,f).

\begin{figure} 
\includegraphics[width=0.99\hsize]{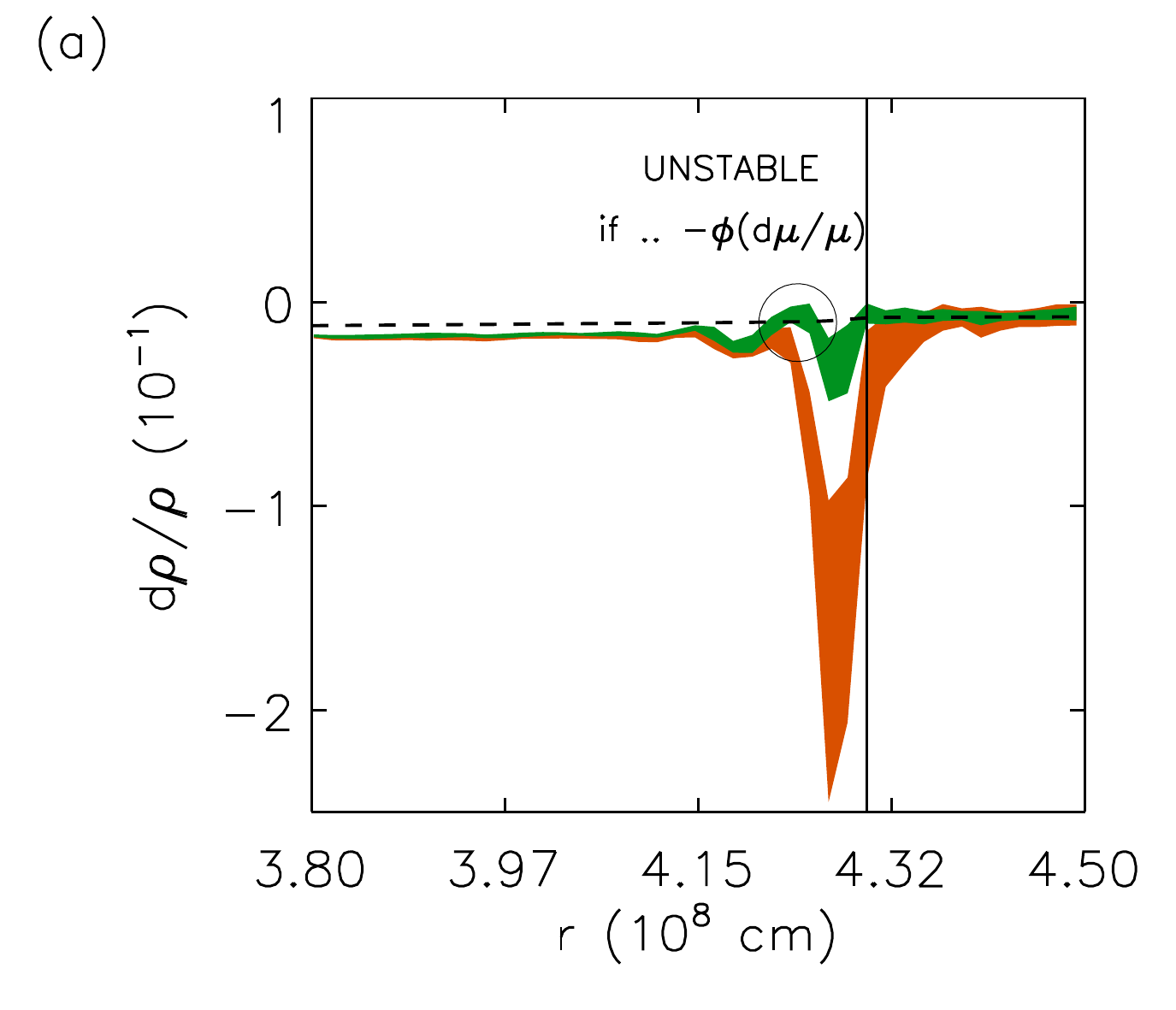}
\includegraphics[width=0.99\hsize]{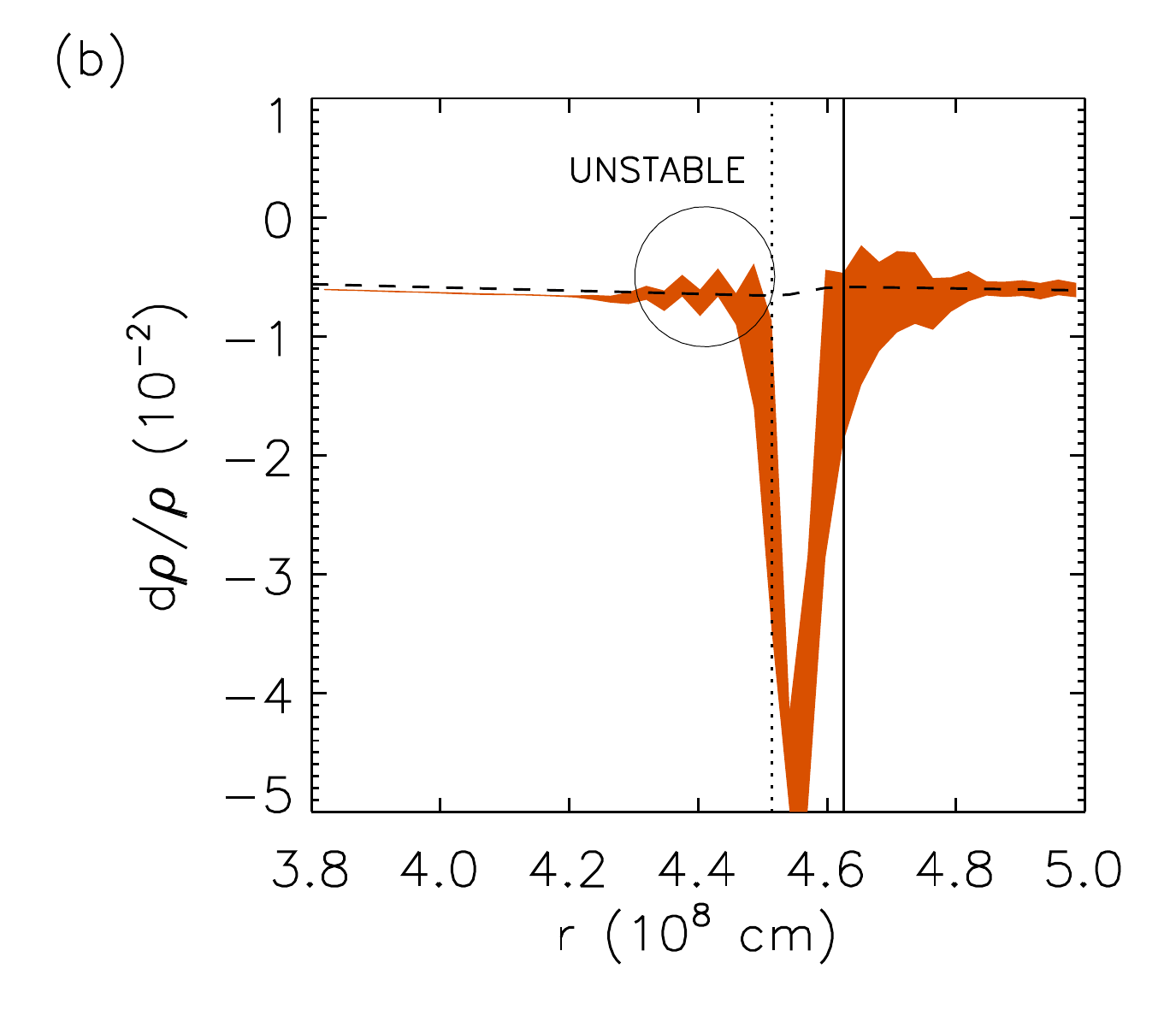}
\caption{Angular averaged relative density fluctuations for adiabatic
  displacements (dashed line) of a fluid element in the 2D oxygen
  shell burning model oxfl.2d as a function of radius at $t= 847\,$s
  (upper panel), and in the 2D core helium flash model hefl.2d.3 at
  $t= 63430\,$s (lower panel). The red shaded area shows relative
  density fluctuations given by $d\rho / \rho = \alpha dP/P - \delta
  dT/T + \phi d\mu/\mu$ (see Eq.\,\ref{eq.drhorho}), while the green
  shaded area gives the same quantity, however assuming a
  $\mu$-gradient of the same absolute size but of opposite sign. The
  width of the shaded areas indicates the variation of the relative
  density fluctuations with polar angle $\theta$ at a given
  radius. The vertical solid line marks the bottom of the convection
  zone (location of $T_{max}$), while the dotted line gives the
  location from where RTFI mixing is launched. }
\label{fig.densover}
\end{figure}

As mentioned before RTFI mixing is similar to mixing by convection in
stars characterized by hot blobs except that it is caused by
sinking cold dense blobs enriched with gas of larger mean molecular
weight. However, there exists another important difference between
both dynamic mixing processes. In convection the surrounding gas has
to have a superadiabatic temperature gradient in order for convection
to set in, while RTFI mixing sets in even if the correspoding layers 
are initially dynamically stable and Ledoux instability 
criterion is not fullfilled. In the latter
case becomes the stability broken only subsequently in our simulations,
when convection boundary becomes partially enriched by material with
higher $\mu$ from convection zone and its enriched dense blobs start to sink.

As it seems that, the $\mu$-gradient is the reason for the observed 
instability,
we analyzed in more detail the dependence of $d\rho / \rho$ on $d\mu/
\mu$ near the boundary of the convection zone. Using the equation of
state $\rho = \rho(P,\, T,\, \mu$), we rewrite the relative (radial)
variation of the density as
\begin{equation}
  \frac{d\rho}{\rho} = \alpha \frac{dP}{P} - \delta \frac{dT}{T} +
                       \phi \frac{d\mu}{\mu} \, ,
\label{eq.drhorho}
\end{equation}

\noindent
where the partial derivative $\alpha = \partial \ln \rho \,/\,
\partial \ln P$ is taken at constant $T$ and ~$\mu$. Layers are stable
as long as their density gradient $(d\rho/ \rho)_{sur}$ does not
exceed the density change experienced by a displaced fluid element
$(d\rho/ \rho)_{ele}$ (Eq.\,\ref{eq.densled}), which we assumed to be
equal to the adiabatic gradient $(d\rho/ \rho)_{ad}$.

The radial distributions of $(d\rho/ \rho)_{sur}$ and $(d\rho/
\rho)_{ad}$ are shown in Fig.\,\ref{fig.densover}(a) for our 2D oxygen
shell burning model oxfl.2d near the base of the convection zone . We
analyzed the stability of these layers by calculating the total
differential $d\rho/ \rho$ (Eq.\,\ref{eq.drhorho}), which depends on
the pressure, temperature, and composition gradients ($dP/P$, $dT/T$
and $d\mu /\mu$, respectively). Given the positive $\mu$-gradient of
the model, we find that $(d\rho/ \rho)_{sur}$ is always smaller than
$(d\rho/ \rho)_{ad}$ at the boundary of the convection zone implying
dynamic stability. Assuming for the composition gradient the same
absolute value, but the opposite sign, we find that $(d\rho/
\rho)_{sur}$ exceeds $(d\rho/ \rho)_{ad}$ below the lower edge of the
convection zone, \ie the corresponding stellar layers at $r \approx
4.2 \times ~10^8\,$cm become unstable.

The positive $\mu$-gradient therefore reduces $(d\rho/ \rho)_{sur}$,
making the layer more stable.  The effect of a negative $\mu$-gradient
is exactly opposite leading to an increase of $(d\rho /\rho)_{sur}$
and a less stable situation compared to that with a positive
$\mu$-gradient.

The destabilizing effect at the base of our core helium flash model is
rather small ($\sim 1\%$ ) due to the smallness of the $\mu$-gradient,
but apparently strong enough to instigate RTFI mixing. This can be
infered from Fig.\,\ref{fig.densover}(b), where $(d\rho/ \rho)_{sur}$
exceeds its adiabatic counterpart below the convection zone for
$4.35\times 10^8\,\mathrm{cm} < r < 4.5\times 10^8\,$cm.

Hence, the driving element behind RTFI mixing is a negative $\mu$-gradient.

\begin{figure} 
\includegraphics[width=0.95\hsize]{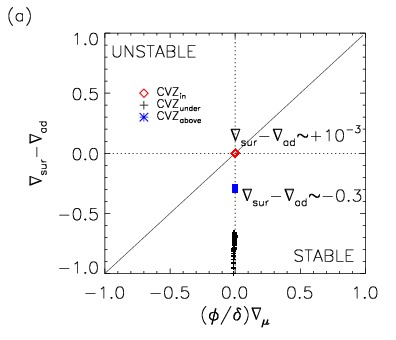}
\includegraphics[width=0.95\hsize]{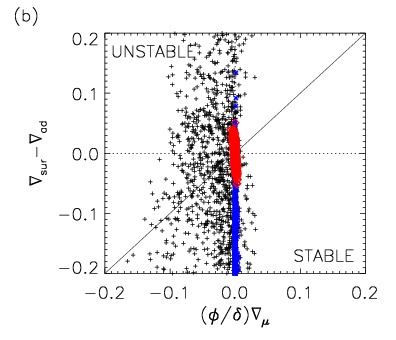}
\caption{Distribution of zones from the 2D model hefl.2d.3 in a
  temperature ($\nabla_{sur} -\nabla_{ad}$) $vs$ composition gradient $(\phi
  /\delta)\nabla_\mu$ plane, where each computational zone is
  represented by a symbol. Zones inside, above and below the
  convection zone are marked by red diamonds, blue asterices, and
  black crosses, respectively. The upper panel (a) gives the distribution
  of zones at $t = 123\,$s, and the lower one (b) at $t = 63430\,$s,
  respectively. The flow in the zones above (below) the diagonal line
  is Ledoux unstable (stable). }
\label{fig.nablasplan}
\end{figure}

\section{Discussion}
\label{sect:disc}

Despite the observed existence of dynamic mixing in stars driven by
chemical gradients, its inclusion into stellar evolutionary
calculations imply non existing observational evidence.  Motivated by
mixing driven by the $^3$He inversion found in hydrodynamic
simulations of hydrogen layers in low-mass red giants by
\citet{Eggleton2006}, \citet{Denissenkov2008} discovered that the time
which stars spend in the bump while climbing the red-giant branch is
significantly prolonged.

These finding suggests that layers with a negative gradient of the
mean molecular weight should not become unstable on dynamical
timescales. However, we observe just the opposite in our hydrodynamic
simulations, where $\nabla_{sur}-\nabla_{ad}-\frac{\phi}{\delta}
\nabla_\mu > 0$ is a very common situation, as demonstrated by
Fig.\,\ref{fig.nablasplan}. It depicts the stability properties of
individual grid zones in the temperature ($\nabla_{sur} -\nabla_{ad}$) $vs$
composition gradient $\frac{\phi}{\delta}\nabla_\mu$ plane.

Initially, all grid zones above and below the convection zone have a
sub-adiabatic temperature gradient ($\nabla_{sur}-\nabla_{ad}<0$) and thus 
are stable, while for grid
zones in the convective shell the temperature gradient exceeds the
adiabatic one by 10$^{-3}$ rendering them unstable
(Fig.\,\ref{fig.nablasplan}\,a). After the onset of convection
the situation changes considerably (Fig.\,\ref{fig.nablasplan}\,b).
Whereas the temperature gradient remains super-adiabatic in many grid
zones inside the convection zone, some of these zones become
sub-adiabatic and their composition changes. This can be understood by
realizing that convection consists of rising hot plumes
(super-adiabatic) of gas enriched by nuclear ash from nuclear burning
and sinking cold plumes (sub-adiabatic) mixed with gas from layers
further out. Chemical species are mixed quite uniformly throughout the
evolved convection zone,\ie the corresponding grid zones tend to gather 
at $\nabla_\mu \sim 0$.  On the other hand, some zones from below the 
base of the convection zone become dynamically unstable as their
 $\nabla_{sur}-\nabla_{ad}-\frac{\phi}{\delta}\nabla_\mu$ is positive.  
These are the zones which participate in RTFI mixing.

\section{Summary}
\label{sect:sum}

We have performed multidimensional hydrodynamic simulations of shell
convection based on three different initial models. These correspond
to helium core at the peak of the core helium flash in a low-mass
star, carbon core during the core carbon flash and oxygen burning
shell above silicon-sulfur rich core of a massive star.

We find a mixing process under the base of convection zones
manifesting itself as over-dense blobs sinking in the direction of
gravity and originating in the convection boundary, which lead to
appearance of overdense fingers enriched by gas of higher mean
molecular weight. This mixing is likely a result of the
Rayleigh-Taylor instability operating due to the presence of negative
mean molecular weight gradient ($\nabla_\mu < 0$) . This is mainly
confirmed by approximate match between theoretical velocity estimate
for an element of matter with excess density over the average of its
surrounding and observed velocities in our 2D simulation with highest
resolution. The theoretical velocities based on observed amplitude of
density fluctuations in the fingers $\Delta \rho/ \rho \sim 5 \times
10^{-4}$ give us a value of $v \sim 7 \times 10^5$\cms, which is close
to the observed velocities reaching values $1 \times 10^{5}\cms <$ v
$< 3.5 \times 10^{5}$\cms.  The fingers are denser and colder than
their surrounding, which rules out their possible explanation as the
thermohaline mixing. The observed fingers lead to a mixing of material
with higher mean molecular weight from convection boundary partially
enriched by gas from above-lying convection zone deeper to stellar
cores and we call it Rayleigh-Taylor finger instability mixing or the
RTFI mixing. The sinking over-dense blobs which cause the fingers are
literally shooting from the convection boundary downwards. The number
of such fingers appearing at given time is correlated with maximum of
square of Brunt-V\"ais\"al\"a frequency in the convective boundary, as
it is proportional to the amplitude of density fluctuations.  If the
regarded amplitude of N$^2$ is higher in the convection boundary, the
fluctuations there are stronger (heavier) too. This in turn increases
the likelihood of having heavy blobs of gas which can sink and create
the fingers. The frequency at which the fingers appear is not clearly
correlated with above-lying convection. However, preferred spots from
where the fingers start are correlated with temperature maxima just
above the convection boundary, which one can associate with up-flows
in the convection zone.

We find similar mixing below the base of the convection zone in the
carbon core during the core carbon flash, too. However, we did not
observe it below the base of the convection zone in our model of shell
convection in oxygen burning shell. The only qualitative difference of
the oxygen burning shell model compared to the others is a positive
mean molecular weight gradient $\nabla_\mu$ (\ie $\mu$ increasing in
direction of gravity) at the bottom of its convection zone.  Hence, we
conclude that the condition $\nabla_\mu < 0$ must be the driving
element of RTFI mixing. We could confirm this also analytically for
the oxygen shell burning model by examining the dependence of density
fluctuations on the $\mu$-gradient at the lower convection zone
boundary.

Implications of RTFI mixing for stellar evolution have still to be
explored. While it almost certainly will not influence the core helium
flash considerably, it could, for instance, have some important
influence on the properties of the deflagration flame during the core
carbon flash.

%
\begin{acknowledgements}
The simulations were performed at the computer center of the
Max-Planck-Society in Garching (RZG). Miroslav\,Moc\'ak acknowledges
financial support from the Communaut\'e francaise de Belgique -
Actions de Recherche Concert\'ees, and from the Institut d'Astronomie
et d'Astrophysique at the Universit\'e Libre de Bruxelles (ULB). The
authors thank Lionel Siess and Casey Meakin for providing us with
initial stellar models. We also want to thank Christophe Almarcha and
Anne De Wit for enlightning discussions on chemo-hydrodynamics, and
for supplying Figure\,1 to us. Finally, we express our gratitude to
Lionel Siess, Rob Izzard, and Casey Meakin for valuable discussions,
and for helpful comments on the manuscript.
\end{acknowledgements}

\bibliography{referenc}

\end{document}